\documentclass[journal]{IEEEtran}
\usepackage[pdftex]{graphicx}
\usepackage[cmex10]{amsmath}
\usepackage{amssymb}
\usepackage{multicol}
\usepackage{colortbl}%
  
\usepackage{longtable}
\usepackage{caption}
\usepackage{subcaption}
\usepackage{cite}
\usepackage{psfrag}
\usepackage{epstopdf}
\usepackage{algorithm}
\usepackage{algorithmic}
\usepackage{widetext}
\usepackage{float}
\usepackage{color}
\usepackage{soul}
\usepackage{upgreek}
\usepackage{multirow}
\usepackage[justification=centerlast]{caption}
\usepackage{tikz,pgfplots}
\usepackage{dblfloatfix}
\usepackage{textcomp}
\usepackage{adjustbox}
\DeclareUnicodeCharacter{2005}{\hspace{0.01em}}
\usepackage{tikz} 
\usetikzlibrary{calc}
\usepackage{mathtools}
\usetikzlibrary{arrows}
\pgfplotsset{compat=newest}
\usepgfplotslibrary{fillbetween}
\usetikzlibrary{patterns}

\usepackage[hyphens]{url}
\usepackage[hidelinks]{hyperref}
\hypersetup{breaklinks=true}
\usepackage{cite}

\definecolor{myBlue}{RGB}{72,125,215}
\definecolor{myOrange}{RGB}{118,54,45}
\definecolor{InfinBlue}{RGB}{72,72,51}

\ifCLASSINFOpdf
\else
\fi
\begin{document}
%
\title{Computational Complexity Optimization of Neural Network-Based  Equalizers in Digital Signal Processing: A Comprehensive Approach}
     \pgfplotsset{
        compat=1.3, 
        my axis style/.style={
            every axis plot post/.style={/pgf/number format/fixed},
            ybar=5pt,
            bar width=8pt,
            x=1.7cm,
            axis on top,
            enlarge x limits=0.1,
            symbolic x coords={MLP, biLSTM, ESN, CNN+MLP, CNN+biLSTM, DBP},
            visualization depends on=rawy\as\rawy, 
            nodes near coords={%
                \pgfmathprintnumber[precision=2]{\rawy}
            },
            every node near coord/.append style={rotate=90, anchor=west},
            tick label style={font=\footnotesize},
            xtick distance=1,
        },
    }
%

\author{Pedro J. Freire, Sasipim Srivallapanondh, Bernhard Spinnler, Antonio Napoli, Nelson Costa, Jaroslaw E. Prilepsky, Sergei K. Turitsyn
\thanks{This paper was supported by the EU  Horizon 2020 program under the Marie Sklodowska-Curie grant agreement 813144 (REAL-NET) and 956713 (MENTOR). JEP is supported by Leverhulme Trust, Grant No. RP-2018-063. SKT acknowledges support of the EPSRC project TRANSNET. B. Spinnler, A. Napoli, and N. Costa acknowledge support from the European Union’ Horizon Europe research and innovation programme, GA No. 101092766 (ALLEGRO) for funding their research.}
\thanks{Pedro J. Freire, Sasipim Srivallapanondh, Jaroslaw E. Prilepsky  and Sergei K. Turitsyn are with Aston Institute of Photonic Technologies, Aston University, United Kingdom, p.freiredecarvalhosourza@aston.ac.uk.}
\thanks{Antonio Napoli and  Bernhard Spinnler are with Infinera R\&D, Sankt-Martin-Str. 76, 81541, Munich, Germany. anapoli@infinera.com.}
\thanks{Nelson Costa is with Infinera Unipessoal, Lda, Rua da Garagem nº1, 2790-078 Carnaxide, Portugal, ncosta@infinera.com.}

\thanks{Manuscript received xxx 19, zzz; revised January 11, yyy.}}

%
%

\markboth{Journal of Lightwave technology , ~Vol.~y, No.~x, November~2023}%
{Shell \MakeLowercase{\textit{et al.}}:  xxxxxxxxxxxx}
%



\maketitle
\begin{abstract}
Experimental results based on offline processing reported at optical conferences increasingly rely on neural network-based equalizers for accurate data recovery. However, achieving low-complexity implementations that are efficient for real-time digital signal processing remains a challenge. This paper addresses this critical need by proposing a systematic approach to designing and evaluating low-complexity neural network equalizers.
Our approach focuses on three key phases: training, inference, and hardware synthesis. We provide a comprehensive review of existing methods for reducing complexity in each phase, enabling informed choices during design.
For the training and inference phases, we introduce a novel methodology for quantifying complexity. This includes new metrics that bridge software-to-hardware considerations, revealing the relationship between complexity and specific neural network architectures and hyperparameters. We guide the calculation of these metrics for both feed-forward and recurrent layers, highlighting the appropriate choice depending on the application's focus (software or hardware).
Finally, to demonstrate the practical benefits of our approach, we showcase how the computational complexity of neural network equalizers can be significantly reduced and measured for both teacher (biLSTM+CNN) and student (1D-CNN) architectures in different scenarios.
This work aims to standardize the estimation and optimization of computational complexity for neural networks applied to real-time digital signal processing, paving the way for more efficient and deployable optical communication systems.




\end{abstract}


%
\IEEEpeerreviewmaketitle
\section{Introduction}

\IEEEPARstart{O}{ver} the last few decades, neural networks (NNs) have begun to find widespread usage in a wide range of signal processing applications: filtering, parameter estimation, signal detection, system identification, pattern recognition, signal reconstruction, time series analysis, signal compression, signal transmission, etc. \cite{miller1992review, feldkamp1998signal,kiranyaz20211d,overview2021}. Audio, video, image, communication, geophysical, and radar scanning data, are examples of important signal types that typically undergo various forms of signal processing \cite{amari1998adaptive, burse2010channel, kahrs1998applications}.
The key capabilities of NNs in signal processing are: performing distributed processing, emulating nonlinear transformations and processes, self-organizing, and enabling high-speed processing communication applications\cite{Govil2000, lusch2018deep, Huttunen2019}. With these properties, NNs can provide a very powerful means of solving many signal processing tasks, particularly in the areas related to nonlinear signal processing, real-time signal processing, adaptive signal processing, and blind signal processing \cite{6975096,amari1998adaptive, luo1998applied,ibnkahla2000applications}.

NN methods have also proven to be efficient in several applications in optical communications, particularly in channel equalization\cite{7359099}. NN structures have demonstrated the potential to significantly enhance transmission quality in various 
scenarios\cite{freire2023reducing}, with computational complexity comparable or better than that of classical approaches\cite{schadler2021soft,napoli2014reduced}. However, additional work is still required to reach product-level applications. One of the most pressing challenges is to demonstrate how such NN-based equalizers can be effectively implemented, taking into account both the training and inference phases. In order to evaluate the computational complexity, the metrics for an assessment should be appropriately addressed.

Real-time signal processing, as an example, is a field that enables technological breakthroughs by effectively incorporating signal processing in hardware: real-time and onboard signal processing are the keys to the evolution of phones and watches into smartphones/smartwatches. To the best of our knowledge, one of the first real-time applications of NNs was discussed in 1989 ~\cite{malkoff1989neural}, and numerous works since then have deliberated the challenges of implementing such solutions in hardware exploiting the notion of computational complexity \cite{sze2017efficient, gysel2016hardware,li2017reducing,yang2017designing, balcazar1997computational, van2020bayesian, baskin2021uniq, deligiannidis2021performance, sidelnikov2018equalization, freire2021performance}. Similarly, in the NN-based equalizer, computational complexity analysis is necessary. 

\begin{figure*}[th]
\centering
\includegraphics[width=0.95\linewidth]{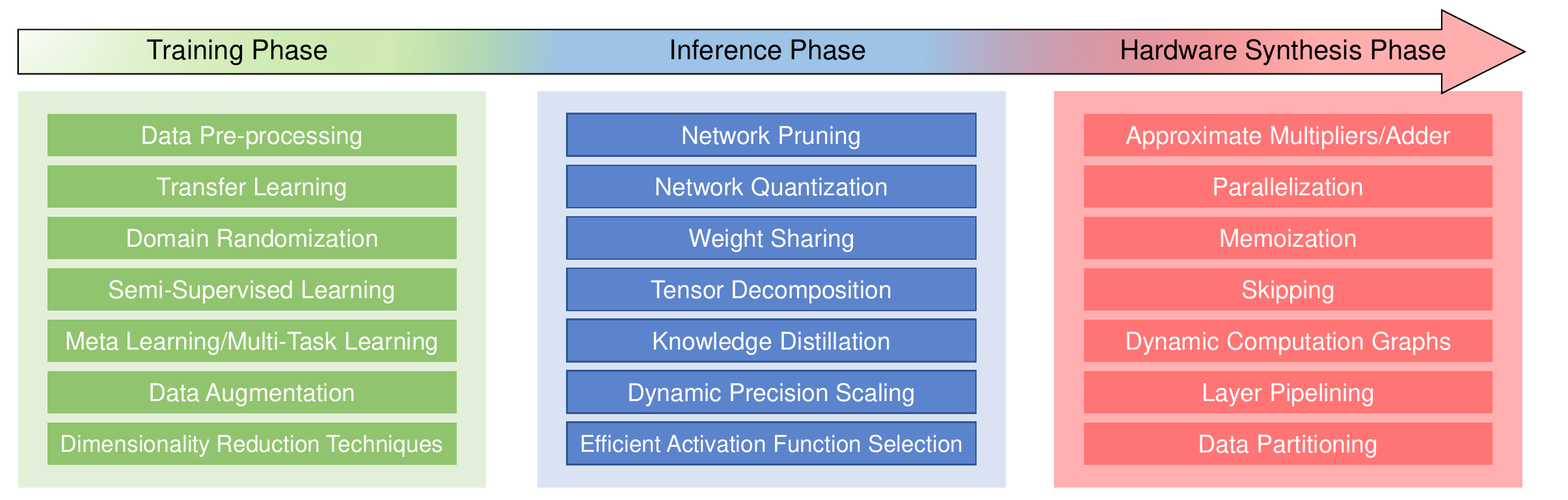}
\caption{Main strategies to design low complex NN-equalizers in training, inference, and hardware synthesis phases.}
\label{fig:overview}
\vspace{-1.5mm}
\end{figure*}

From a computer science perspective, computational complexity analysis is almost always attributed to the Big-$O$ notation of the algorithm \cite{maass1994computational,alizadeh2020managing,wiedemann2019compact}. In general, the Big-$O$ notation is used to express an algorithm's complexity while assessing its efficiency, which means that we are interested in how effectively the algorithm scales with the size of the dataset in terms of running time \cite{amin2008analysis,kerr2005big, blondel2000survey}. However, from the engineering standpoint, the Big-$O$ is often an oversimplified measure that cannot be immediately translated into the hardware resources required to realize the algorithm (NNs) in a hardware platform \cite{gysel2016hardware}.

Due to this problem that refers to the absence of some ``universal'' measure, various works started to present complexity in terms of multiply and accumulate (MAC) \cite{sze2017efficient, gysel2016hardware,li2017reducing,yang2017designing}, Kolmogorov complexity \cite{balcazar1997computational}, the number of bit-operations (BOP) \cite{van2020bayesian, baskin2021uniq}, the number of real multiplications (RM) \cite{deligiannidis2021performance, sidelnikov2018equalization, freire2021performance}. However, it is not always clear when to use each specific metric, and, more importantly, none of the metrics mentioned above show the benefits of using different strategies of quantization to reduce the complexity of implementing the multipliers.

As far as we know, no work has so far unified the computational metrics itemized above such that we have no universal metrics to compare the complexity when different types of quantization are applied to NN structures. In this paper, we solve this issue by carrying out a systematic computational complexity analysis for a zoo of NN layer types. In addition, we introduce a new useful metric: we coined `the number of additions and bit shifts' (NABS). This metric takes into account the impact of the weights' quantization type on the reduction of the multipliers' implementation complexity used in an NN layer. Overall, we intend our work to give largely universal measures of complexity to establish a comparison baseline depending on whether the application is software- or hardware-based. 

The paper is organized as follows. Firstly, Sec.~\ref{Sec:low-complex-methods} provides an overview of the main strategies enabling low computational complexity NN-based equalizers from training to hardware synthesis, while still maintaining attractive Q-factor gains.
In Sec.~\ref{Sec:metrics} we describe the details of different computational metrics for the training and inference stages. Sec.~\ref{Sec:complexity} outlines a method for computing the computational complexity of diverse neural network layers based on their hyperparameters, traversing from the software to the hardware level. 
The computational complexity growth against the design parameters of each neural network layer is discussed in Section~\ref{Sec:results}. Additionally, in Section~\ref{Sec:case-study}, we explore the impact of quantization on different computational complexity metrics and present a practical study in the realm of channel equalization. This study exemplifies complexity reduction approaches and elucidates the performance and complexity trade-offs associated with compensating nonlinearities in optical coherent transmissions.
Our findings are summarized in the conclusion.

\section{Overview on Complexity Reduction In Neural Network Equalizers}\label{Sec:low-complex-methods}
To achieve optimal computational complexity in the pursuit of efficient NN equalizers for resource-constrained hardware, it is necessary to thoroughly investigate three crucial phases: training, inference, and hardware synthesis phases. Fig.~\ref{fig:overview} illustrates the most common techniques applied to reduce the complexity of NN-based equalizers in each of these phases. 

\subsection{Complexity reduction in training}
First, during training, reducing complexity is crucial for the efficient and practical deployment of hardware with limited resources. Various complexity reduction techniques are employed to enhance efficiency and performance. Techniques such as transfer learning or approaches to improve generalization, such as data augmentation, domain randomization, and semi-supervised learning, can be applied. These approaches indirectly reduce the need for large amounts of original training data. Effective generalization reduces the need for complex models with a high number of parameters, resulting in faster and more efficient training.

\textbf{Data pre-processing} is a crucial step in preparing the input for the NN training. This can be a main factor in enhancing the model efficiency and accelerating convergence because high-quality input data contributes to stable training, improves generalizability, and leads to successful data interpretation by the model\cite{nawi2013effect}. Data pre-processing involves data normalization to have the features on a consistent scale, or feature engineering, which selects and transforms the input features to emphasize the relevant information and get rid of the noisy, irrelevant data.

\textbf{Transfer learning (TL)} adapts the knowledge acquired in the source tasks to the related target tasks. This technique can considerably reduce the training time and resources required. TL is particularly useful when training the models on limited computational resources, as it allows the NN to inherit knowledge from a larger pre-trained model and fine-tune it for a specific task. TL was investigated in equalization tasks in both directly\cite{xu2020transfer} and coherently\cite{freire2021transfer} detected systems. Ref.~\cite{freire2021transfer} demonstrated the potential of TL to reduce the number of training epochs and the training dataset without impacting the equalizer’s performance.

\textbf{Domain randomization} is a systematic approach for data generation aiming to improve the generalization and the robustness of machine learning models in new environments~\cite{tobin2017domain}. Domain randomization generates training data from a random distribution with given desired properties and stores it in a library accessible by NN. By using synthetic data, this approach reduces the dependence on real-world data, thus improving the training efficiency ~\cite{freire2022domain}. This technique is especially valuable when dealing with complex and dynamic environments as the model becomes more adaptable to variations encountered during deployment.

\textbf{Semi-supervised learning} combines labeled and unlabeled data in training, allowing the model to learn from both. Semi-supervised learning enables NN to leverage the available labeled data more effectively by incorporating information from the unlabeled samples. This method enhances the model's performance without requiring additional labeled data, which makes the model more flexible to transmission changes.  This method resembles decision-directed adaptive equalization\cite{zhou2020adann} for channel equalization.

\textbf{Meta learning} involves training models that can efficiently adapt to new tasks with only a small amount of training data based on experiences gained from a variety of learning tasks \cite{finn2017model}. This approach explicitly trains the model parameters to enable efficient generalization on new tasks with a small number of gradient steps and minimal training data, making it simple to fine-tune. 

\textbf{Multi-Task Learning} is a paradigm where a single model is trained to perform multiple but related tasks simultaneously. In contrast, traditional single-task learning trains multiple separate models for each task independently. Multi-task learning leverages shared representations across tasks and the model's parameters are optimized jointly across all tasks. This training approach can lead to better generalization of the model and reduce the need to deploy several models for different tasks and does not require re-training when performing related tasks. This technique not only reduces the number of models that need to be trained but also reduces the complexity in the inference phase. However, this technique can exhibit a trade-off between overall performance and specific task performance. This approach has been shown to be efficient in the NN-based equalizers in both IM/DD \cite{liu2023area} and coherent systems \cite{srivallapanondh2023multi}. 

\textbf{Data augmentation} allows datasets to be more diverse and representative by artificially generating additional data points from existing data\cite{alomar2023data}. Data augmentation used in optical NN-based equalizers is a technique to improve equalization performance and decrease the training complexity of supervised learning in nonlinearity mitigation. In supervised learning tasks, normally a large training dataset is required. The model will also need to be re-trained when the channel conditions change. However, big data collection can be challenging.  The efficient use of a limited dataset is more desirable for practical implementation. Ref.~\cite{9333417} showed that data augmentation reduces the size of the dataset up to 6 times while maintaining the optical performance. This technique enables a less overfitting model, fewer model parameter requirements, and a faster convergence of the training. 

\textbf{Dimensionality reduction techniques} address the challenges associated with high-dimensional input spaces. These techniques aim to capture and retain the most informative aspects of the data while reducing the number of input features significantly\cite{velliangiri2019review}.  Principal Component Analysis (PCA) \cite{mackiewicz1993principal}, for instance, transforms the original features into a lower-dimensional space defined by principal components (a new set of uncorrelated variables), retaining the maximum variance.

\subsection{Complexity reduction in inference}
Next, during inference, the NN must accurately equalize the input signal using the minimum computational resources while meeting the required performance metrics. As in the real world, the environments and the resources are more constrained. This result can be achieved by using different techniques, for example, network pruning, sparse representation, knowledge distillation (KD), and tensor decomposition.

\textbf{Network pruning} reduces the complexity of NNs by removing redundant or less significant parameters (weights, connections,  neurons, or layers) from a trained NN. Pruning can be carried out without significantly affecting equalization performance, as described in~\cite{freire2023reducing,ge2020compressed, ron2022experimental}. By eliminating unnecessary parameters, this approach reduces the model's size and computational requirements during inference. Sparse connectivity, achieved through pruning, enables faster execution of the NN on hardware. In most cases, pruning in optical channel equalization has been restricted to the feedforward NN, however, Ref.~\cite{freire2023reducing} extended the investigation to the case of recurrent equalizer in coherent optical transmission.

\textbf{Network quantization} reduces the precision of the weights and activation functions in NN, for instance, 32-bit floating-point values are converted to 8-bit integers. However, the trade-off between complexity and performance should also be carefully considered, because there might be a performance sacrifice when precision is drastically reduced. This approach has the potential to reduce complexity and memory usage, aiming to make the model more efficient for deployment on resource-constrained device~\cite{freire2023reducing, koike2021zero}.

\textbf{Weight Sharing} compresses the NN weights and biases by keeping only a small number of non-zero coefficients. The redundant or similar values of weights in the NN are collapsed into a single shared weight \cite{freire2023reducing}. The reduction in the number of unique values of weights decreases the number of parameters that need to be computed and stored, leading to a more compact and memory-efficient model.

\textbf{Tensor decomposition} decomposes high-dimensional data into a lower-dimensional space \cite{kolda2009tensor}. In other words, a multidimensional tensor is broken down into a combination of simpler tensors. By decomposing tensors, especially weight tensors in NN, into smaller and more manageable components, tensor decomposition reduces the number of parameters and computations needed in the inference phase. In \cite{9125675}, the authors showed that the sparse decomposition of the tensor in convolutional filters can successfully reduce model complexity and memory usage during inference.

\textbf{Knowledge Distillation (KD)} is applied to transfer knowledge from a larger model (teacher) to a more compact one (student) using teacher predictions to assist student learning. KD can reduce the size of the model\cite{gautam2022optidistillnet}.  The distilled model retains the essential information from the teacher model, making it suitable for deployment in resource-constrained environments during the inference process. Ref.~\cite{srivallapanondh2022knowledge} proved to use KD to accelerate the inference of the NN equalizer by recasting the RNN-based equalizer into a feed-forward-based equalizer without significantly compromising the equalization performance.

\textbf{Dynamic precision scaling (DPS)} adjusts the precision of the numerical values of the weights and during computation dynamically, based on the specific requirements of each computation.  With this approach, the NN can utilize lower precision when the accuracy demands allow, as DPS optimizes the utilization of available resources. This approach provides an effective reduction of complexity during inference. Ref.~\cite{taras2018quantization} showed that DPS could be used in both the forward pass (inference) and backward pass for training. 

\textbf{Efficient Activation Function Selection} can play an important role in complexity reduction. The expensive activation functions, e.g. hyperbolic tangent or sigmoid can be replaced by the approximated alternatives or with the Look-up Table to reduce the computation \cite{freire2023implementing}. Simpler functions such as ReLU (Rectified Linear Unit) are also commonly chosen because of their simplicity in calculation and speed.

\subsection{Complexity reduction in hardware synthesis}
Complexity reduction techniques of NN in hardware synthesis play a crucial role in optimizing the NN implementation on dedicated hardware. In hardware synthesis, the NN is mapped onto the hardware architecture, and the hardware design is optimized to achieve the desired performance while minimizing resource utilization. There are some techniques to reduce the complexity of the hardware, such as multiplier/adder approximations, parallelization,  memoization, and skipping. The choice of suitable techniques depends on the NN architecture, the characteristics of the target hardware, and the desired trade-off between computational efficiency and model accuracy.

\textbf{Multiplier/adder approximation} is to reduce the hardware resource requirements by approximating multiplier and adder which are key components of the hardware implementation for NN computation\cite{nojehdeh2020efficient}. 
The approximation replaces full-precision multipliers and adders with less resource-intensive multipliers and adder implementations, such as approximate adders or low-precision. Binary or ternary multipliers are examples of low-precision alternatives. By using lower-precision multipliers and adder implementations, the overall hardware complexity is reduced. This can lead to more efficient use of hardware resources without significantly sacrificing model accuracy.

\textbf{Parallelization} involves dividing the neural network into multiple sub-networks to be processed simultaneously, aiming for faster and more efficient execution in terms of latency and throughput \cite{huang2021recurrent}. This methodology capitalizes on parallel hardware architectures, such as GPUs, yielding heightened computational efficiency. As detailed in Ref. \cite{nichols2021survey}, Parallelization can take various forms including: \textbf{Data Parallelism}, where multiple NN instances operate simultaneously on distinct data batches; \textbf{Model Parallelism}, involving the division of a single neural network across multiple processors or GPUs, with different components processed on separate devices\footnote{In addition to the aforementioned parallelization techniques, another subcategory worth mentioning is intra-layer parallelism, often referred to as Tensor Parallelism. This method entails parallelizing computations within a single layer of the neural network. Specifically, it involves partitioning large tensors, such as weight matrices, of a layer across multiple devices, facilitating parallel computations on these segmented chunks.}; and \textbf{Pipeline Parallelism (Inter-Layer parallelism)}, which segments the neural network computation into stages, each executed by a distinct processing unit such that the data flow resembles a sequential assembly line.

\textbf{Memoization} stores and reuses intermediate results of expensive computations in memory to avoid recalculation when the same input reappears\cite{armeniakos2022hardware, della2015performance}. This can be especially beneficial in RNN or other architectures with repetitive computations, leading to improved hardware efficiency. Even the simple implementation of memoization in Ref.~\cite{della2015performance} could speed up different experiments with different workloads ranging from 7\% up to 25\%.

\begin{figure*}[ht!]
    \centering
\includegraphics[width=0.8\textwidth]{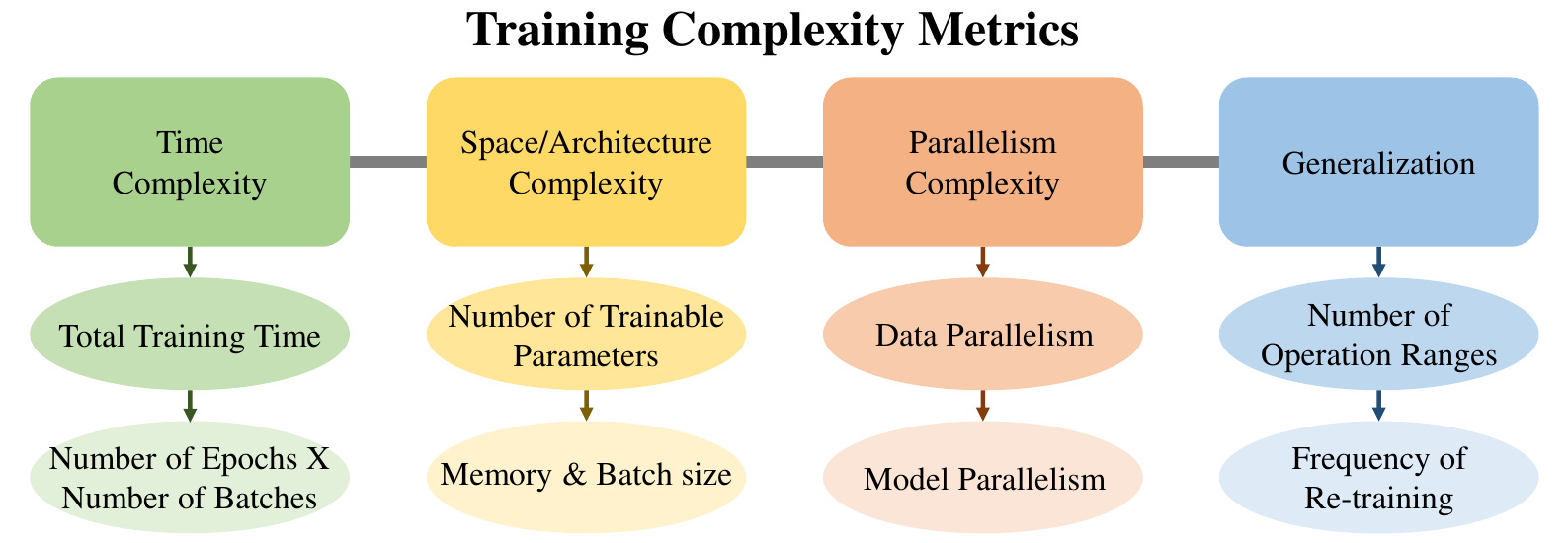}
    \caption{Main metrics to evaluate complexity in the training phase for NN-equalizers.}
    \label{fig:training_complexity_metrics}
\end{figure*}

\textbf{Skipping} can be used to decrease the executed workload and reduce computational costs. This method selectively skips certain computations based on their relevance to the final output or the predefined conditions. Skipping approximations can be performed by a simple calculation to evaluate if a more complex computation can be eliminated \cite{armeniakos2022hardware}.  

\textbf{Dynamic computation graph} allows the structure of the NN to change dynamically at runtime \cite{looks2016deep}. While the static graph fixes the structure of the NN (like the sequence of operations and connections) before the beginning of training, the dynamic graph constructs the structure of the NN on-the-fly during execution. This approach allows flexibility as it adapts the structure efficiently depending on varying input types and conditions. 

\textbf{Layer pipelining} splits the processing of different layers in NN into sequential stages that overlap in time to allow parallelization of the computation. This method \cite{huang2019gpipe} allows scalable model parallelism with high hardware utilization and training stability. Pipelining algorithm library, GPipe, from \cite{huang2019gpipe} is a library to train a giant NN, with efficiency (speeds up the process), flexibility (supports any deep network), and reliability (guarantees consistent training).

\textbf{Data partitioning} is an approach to divide the input data into subsets to be processed in parallel by different hardware components independently \cite{cannas2006data}. After that, the result of each partition is combined. Ref.~\cite{ghis2021mixed} demonstrated that with their data partitioning approach, only small memory storage is required, instead of duplicating the whole data set size over all the processing units.

\subsection{Insights for low-complexity NN equalizers implementation}
For the training, we believe that one of the most important and promising techniques in reducing computational complexity and enhancing the generalization of NN-based equalizers is multi-task learning. As in real-world scenarios, the equalizers require reconfiguration and must be adjustable to compensate for the variation of impairments as the channel characteristics change \cite{srivallapanondh2023multi}. By implementing multi-task learning, the model does not need to be re-trained if performing in the range of knowledge acquired in training, but the model might experience a trade-off between overall performance and specific task performance. This robust approach enables a more practical utilization of NN-based equalizers, aligning closely with real-world demands.

For the inference, we advocate for the utilization of weight clustering techniques to optimize model size and computational complexity. This approach enables us to reduce the NN model's footprint, approaching levels comparable to constrained hardware environments such as CDC complexity \cite{freire2023reducing}. Weight clustering achieves this by consolidating the number of distinct multipliers in matrix multiplication operations to at least the number of clusters per input element. Additionally, it facilitates heterogeneous quantization by minimizing the number of bits required to represent the weights effectively. Such strategies are pivotal in hardware implementations, where resource constraints necessitate efficient utilization of computational resources.

In the domain of hardware synthesis, we investigate various strategies to optimize the implementation of neural network-based equalizers. This involves delving into partitioning schemes, which encompass different strategies for distributing data and computations across processing units. One such approach involves layer-wise partitioning, where each layer of the neural network is allocated to specific processing units. Alternatively, channel-wise partitioning allocates computations related to individual channels of input data to separate processing units. These partitioning schemes aim to exploit parallelism within the neural network, thereby enhancing hardware efficiency and performance.

Furthermore, we explore parallelization architectures tailored for the efficient execution of parallel processing tasks. Systolic arrays represent one such architecture that orchestrates computations through a pipeline of processing elements arranged in a grid-like fashion. Specialized neural network accelerators are also investigated, leveraging dedicated hardware components optimized for executing neural network operations in parallel. These architectures are designed to exploit inherent parallelism within neural network computations, leading to significant improvements in performance and throughput.

\begin{figure*}[ht!]
    \centering
\includegraphics[width=\textwidth]{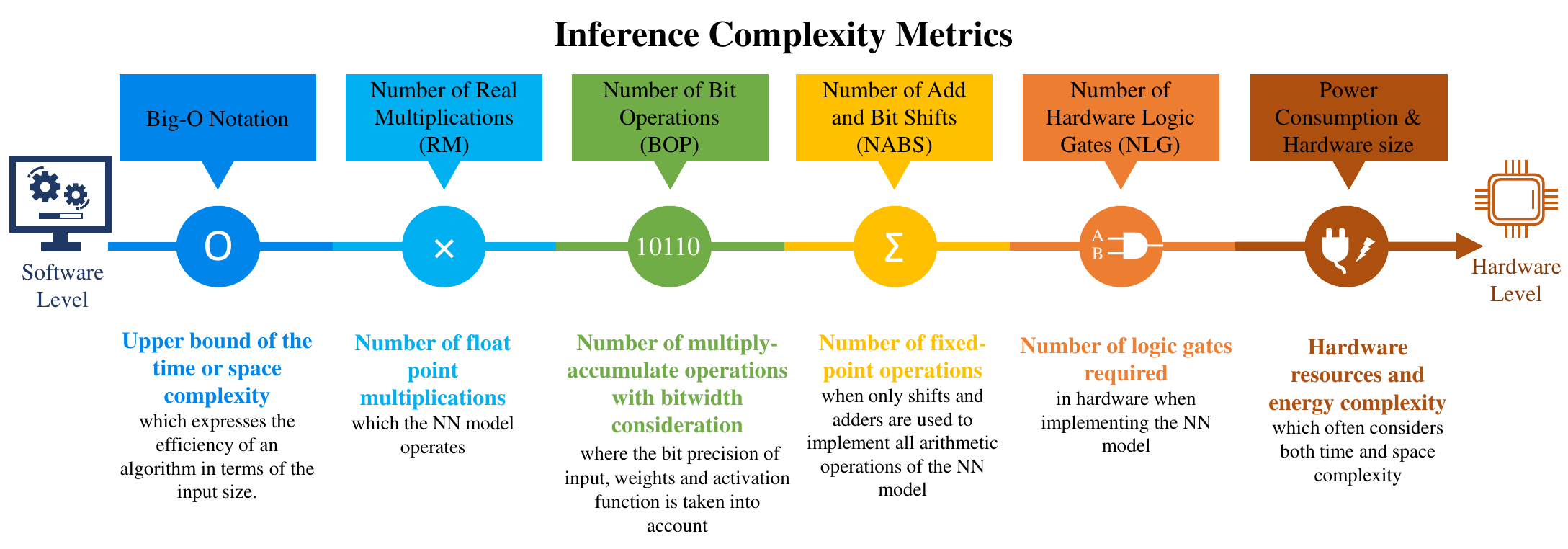}
    \caption{Main metrics to evaluate complexity in the inference phase for NN-equalizers.}
    \label{fig:infer_complexity_metrics}
\end{figure*}

In addition to partitioning and parallelization strategies, we focus on memory access optimization techniques to minimize overhead during data retrieval and processing. This involves optimizing memory access patterns to minimize latency and maximize bandwidth utilization. Techniques such as data prefetching, caching, and memory banking are explored to streamline memory access operations and alleviate bottlenecks associated with partitioned data processing. By optimizing memory access, we aim to enhance overall system efficiency and throughput in hardware implementations of neural network-based equalizers.

\section{Complexity Metrics (Training and Inference)} \label{Sec:metrics}
After implementing the mentioned complexity reduction strategies, it is essential to evaluate their effectiveness. This section gives a comprehensive understanding of the model's complexity during the training and real-time inference phases on the target hardware platform. Fig.~\ref{fig:training_complexity_metrics} and  Fig.~\ref{fig:infer_complexity_metrics} show the metrics used to measure the complexity of training and inference phases, respectively.

\subsection{Complexity metrics for training}
For training, the computational complexity of the NN should be appropriately evaluated, as it allows for efficient resource allocation and is useful for comparing different models to assess the efficiency and effectiveness of various model architectures. Several metrics can be used to assess the complexity of NNs, which can be categorized into four key areas: time, space/architecture complexity, parallelism complexity, and generalization. Adopting a multidimensional perspective is important because a single metric cannot provide a holistic understanding of the true complexity of training. Each dimension offers unique insights, guiding informed trade-offs between model performance, resource requirements, and generalization capabilities. Next, we will further detail each of these key areas of training complexity measurements:

\textbf{Time complexity:} The traditional measures refer to the training time and the number of epochs required to achieve the desired performance. These metrics also take into consideration the learning rate and the 
optimizer used. Even though the training time metric and the number of epochs metrics show, to some degree, the training complexity, these metrics are a poor benchmark since the training time depends heavily on the hardware resources used and the size of the training dataset. In addition, two NNs with the same number of epochs to achieve the same performance can have very distinct training time, also depending on the batch size.  To address these issues, an additional metric is proposed. The product of the number of epochs and the number of batches (NENB), reflects the model's computational demand.  More training epochs generally indicate a more complex and computationally demanding model. The number of batches refers to the number of subsets of data used during each epoch, which is affected proportionally by the dataset size and batch size. The number of epochs and the number of batches cannot be evaluated separately, as one model may require more epochs but fewer batches, while another model may require fewer epochs but more batches. Lastly, FLOPs (Floating Point Operations) can be used to measure the number of floating-point operations required to train the model. A higher number of FLOPs typically indicates higher time complexity.

\textbf{Space/Architecture Complexity:} The number of trainable parameters, while commonly used, may not fully capture the complexity due to different architectural designs. For example,  two NNs with the same number of trainable parameters can have very distinct training complexity~\cite{freire2022neural}. The model architecture indicates the complexity of the NN architecture itself, such as the depth,  width (e.g. the number of layers and neurons) of the network, and specific architectural choices such as recurrent, or convolution. The next metric is the memory requirement for storing the weights, biases of the NN, and intermediate computations during training. The space complexity can be influenced by the batch size used during training. Larger batch sizes might require more memory, particularly on GPU devices.

\textbf{Parallelism Complexity:} This aspect consists of data parallelism and model parallelism. Data parallelism is the parallelization of training across multiple devices by splitting the dataset. Model parallelism is about distributing the model across different devices for processing. Parallelizability also refers to how scalable the training process is with added computational resources and the efficiency of distributing the training process across multiple GPUs. For example, the MLP feed-forward NN is fully parallelizable and can result in faster training. To be more specific, the training time of RNN compared to the MLP with the same number of trainable parameters can be significantly longer, as the recurrent architecture of RNN is more complex than the feed-forward structure of the MLP.

\textbf{Generalization:} The flexibility/generalizability is assessed by estimating the number of operational ranges in which the NN equalizer operates with an acceptable gain. If the NN can only perform a specific task, it requires frequent re-training in the future, contributing to the overall complexity. Therefore, the NN that performs well in different but related tasks without re-training is preferable \cite{srivallapanondh2023multi}.

\subsection{Complexity metrics for inference}
Accurate computational complexity evaluation is critical in the design of digital signal processing (DSP) devices to better understand the implementation feasibility and bottlenecks for each device's structure.
With this in mind, we summarize the four most commonly used criteria for assessing computational complexity, from the software level to the hardware level, in Fig.~\ref{fig:infer_complexity_metrics}.

\subsubsection{Real multiplications}
The first, most software-oriented, level of estimation traditionally deals only with counting the number of real multiplications of the algorithm \cite{jacobsen2007fast,spinnler2010equalizer} (quite often defined per one processed element, say a sample or a symbol). This metric is the number of real multiplications (RM). When comparing computational complexity, the purpose of this high-level metric is to consider only the multipliers required, ignoring additions, because the implementation of the latter in hardware or software is initially considered cheap, while the multiplier is generally the slowest element in the system and consumes the largest chip area \cite{mirzaei2006fpga,jacobsen2007fast}. This ignoring of the additions can also be easily understood by looking at the Big-$O$ analysis of multiplier versus adder. When multiplying two integers with $n$ digits, the computational complexity of the multiplication instance is $O(n^2)$, whereas the addition of the same two numbers has a computational complexity of $\Theta(n)$ \cite{jahani2009zot}\footnote{The Big-$O$ notation represents the worst case or the upper bound of the time required to perform the operation, Big Omega ($\Omega$) shows the best case or the lower bound, whereas the Big Theta ($\Theta$) notation defines the tight bound of the amount of time required; in other words, $f(n)$ is claimed to be $\Theta(g(n))$ if $f(n)$ is $O(g(n))$ and $f(n)$ is $\Omega(g(n))$.}. As a result, if you are dealing with float values with 16 decimal digits, multiplication is by far the most time-consuming part of the implementation procedure.
Therefore, when comparing solutions that use floating-point arithmetic with the same bitwidth precision, the RM metric provides an acceptable comparative estimate to qualitatively assess the complexity against some existing benchmarks (e.g. against the DSP operations for optical channel equalization tasks \cite{spinnler2010equalizer}). 

\subsubsection{Number of bit-operations}
When moving to fixed-point arithmetic, the second metric known as the number of bit-operations (BOP) must be adopted to understand the impact of changing the bitwidth precision on the complexity. The BOP metric provides a good insight into mixed-precision arithmetic performance since we can forecast the BOP needed for fundamental arithmetic operations like addition and multiplication, given the bitwidth of two operands. In a nutshell, the BOP metric aims to generalize floating-point operations (FLOPs) to heterogeneously quantized NNs, as far as the FLOPs cannot be efficiently used to evaluate integer arithmetic operations \cite{baskin2021uniq, hawks2021ps}. For the BOP metric, we have to include the complexity contribution of both multiplications and additions, since now we evaluate the complexity in terms of the most common operations in NNs: the multiply-and-accumulate operations (MACs) \cite{baskin2021uniq, hawks2021ps,wu2018deep}. However, the BOP accounts for the scaling of the number of multipliers with the bitwidth of two operands, and the scaling of the number of adders with the accumulator bitwidth. Note that since most real DSP implementations use dedicated logic macros (e.g. DSP slice in Field Programmable Gate Arrays [FPGA] or MAC in Application Specific Integrated Circuit [ASIC]), the BOP metric fits as a good complexity estimation metric inasmuch as the BOP also accesses the MAC taking into account the particular bitwidth of two operands.

\begin{table}[t]
\caption{Capacity ranges for XC4000 Series CLB Resources given in Ref.~\cite{staff59gate}.}
    \label{tab:CLB_gatecounting1}
    \centering
    \resizebox{0.48\textwidth}{!}{
\begin{tabular}{|c|c|}
\hline
\textbf{CLB Resource} & \textbf{Logic Gate Range} \\ \hline\hline
Gate range per 4-input LUT (2 per CLB) & 1 to 9 \\ \hline
Gate range per 3-input LUT & 1 to 6 \\ \hline
Gate range per flip-flop (2 per CLB) & 6 to 12 \\ \hline
Total gate range per CLB & 15 to 48 \\ \hline
Estimated typical number of gates per CLB & 28.5 \\ \hline
\end{tabular}}
\end{table}

\subsubsection{Number of additions and bit shifts}
 The progress in the development of new advanced NN quantization techniques \cite{li2019additive,Koike2021,elhoushi2021deepshift,you2020shiftaddnet} allowed implementing the fixed point multiplications participating in NNs efficiently, namely with the use of a few bit-shifters and adders \cite{gentili1995efficient, evans1994efficient, lee2003frequency}.  Since the BOP lacks the ability to properly assess the effect of different quantization strategies on the complexity, a new, more sophisticated metric is required there.
 We introduce the third complexity metric that counts the number of total equivalent additions to represent the multiplication operation, called the number of additions and bit shifts (NABS). The number of shift operations can be neglected when calculating the computational complexity because, in the hardware, the shift can be performed without extra costs in constant time with the $O(1)$ complexity. Even though the cost of bit shifts can be ignored due to the aforementioned reasons, and only the total number of adders has to be accounted for to measure the computational complexity, we prefer to keep the full name ``number of additions and bit shifts'' to highlight that the multiplication is now represented as shifts and adders. 

\subsubsection{Number of logic gates}
The metric closer to the hardware level is the number of logic gates (NLG) that is used for our evaluating method's hardware (e.g. ASIC or FPGA) implementation. It is different from the NABS metric, as now the true cost of implementation is to be presented. In this case, in contrast to the other complexity metrics, the cost of activation functions is also taken into account because, to achieve better complexity, they are frequently implemented using look-up tables (LUT) rather than adders and multipliers. Additionally, other metrics like the number of flip-flops (FFs) or registers, the number of logic blocks used for general logic and memory blocks, or other special functional macros used in the design, are also relevant. As it is clear from this explanation, there is no straightforward equation to convert the NABS to the NLG as the latter depends on the circuit design adopted by the developer. Tools such as Synopsys Synthesis \cite{kurup2012logic} for ASIC implementation can provide this kind of information. However, regarding the FPGA design, it is harder to get a correct estimate of the gate count from the report of FPGA tools \cite{li2016efficient}.

In this paper, we advocate that the NLG metric should be applied to count the number of logic gates used to implement the hardware piece, similar to the concept of the Maximum Logic Gates metric for FPGA devices \cite{staff59gate}. The Maximum Logic Gates metric is utilized to approximate the maximum number of gates that can be realized in the FPGA for a design consisting of only logic functions\footnote{On-chip memory capabilities are not factored into this metric.}. Additionally, this metric is based on an estimate of the typical number of usable gates per configurable logic block (CLB) or logic cell multiplied by the total number of such blocks or cells \cite{staff59gate}. Concerning the correspondence between CLB and logic gates number, see Table \ref{tab:CLB_gatecounting1}.

It should be noted that Table \ref{tab:CLB_gatecounting1}  is based on an older, now obsolete, 4-input LUT architecture \cite{staff59gate}. Newer FPGA families now feature a 6-input LUT architecture, and to address the resource consumption for the new generation of devices, a reasonable approximation would be to increase the ‘maximum gate range equivalent per LUT’ figure used in \cite{staff59gate} by 50\%. Note that the gate equivalence figures for FF’s (registers) still hold true for the 6-input architecture.
It is also worth noting that the CLB architecture has changed substantially since Ref.~\cite{staff59gate} was published, such that we include Table \ref{tab:CLB_gatecounting2} linking CLB-gates with a more up-to-date 6-input architecture.

 
\begin{table}[t]
\caption{Estimated capacity ranges for 6 input LUT-based CLB Resources.}
    \label{tab:CLB_gatecounting2}
    \centering
    \resizebox{0.48\textwidth}{!}{
\begin{tabular}{|c|c|}
\hline
\textbf{CLB Resource} & \textbf{Logic Gate Range} \\ \hline\hline
Gate range per 6-input LUT (8 per CLB) & 6 to 15 \\ \hline
Gate range per flip-flop (16 per CLB) & 6 to 12 \\ \hline
Total gate range per CLB & 144 to 312\\ \hline

\end{tabular}}
\end{table}

To conclude, we comment on universal metrics between the FPGA and the ASIC implementations. We emphasize that calculating an ASIC gate equivalent to an FPGA DSP slice is not a straightforward task because not all features are necessarily required when implementing the specific arithmetic function in an ASIC. However, utilizing the estimation approach laid out in Ref.~\cite{staff59gate}, a figure can be obtained.  Using the Xilinx Ultrascale + DSP48E2 slice basic multiplier functionality as an example (see Xilinx UG579 Fig. 1-1 in Ref.~\cite{Ultrascale}) and pipelining it for maximum performance, it is possible to estimate the number of FFs and adders required for such an ASIC equivalence. Taking into account the structure of the multiplication of a $m$-bit number by a $n$-bit number, implemented using an array multiplier architecture, it is equivalent to $m\times n$ AND gates, $n$ half adders, and $(m-2)\times  n$ full adders\footnote{Note that a half adder is equivalent to 1 AND gate + 1 XOR gate, and a full adder is equal to 2 AND gates + 2 XOR gates + 1 OR gate}. For example, the ASIC equivalence of a 27$\times$18 multiplier in an FPGA would have
486 AND gates, 18 half adders, 450 full adders, and 90 FFs.

Note that, estimating the NABS that can be implemented on FPGA hardware presents a considerable challenge, owing to the multitude of factors influencing FPGA implementations. These factors include resource availability such as the number of CLBs and LUTs, the efficacy of programming methodologies to effectively utilize available resources, as well as constraints related to routing, placing, clock frequency, and throughput limitations, among others. Despite the complexity of this estimation, an initial assessment can be made by considering the potential parallelism of NABs per clock cycle, primarily leveraging LUTs and DSP blocks. For instance, one can consider the Xilinx FPGA VCK190 as an example. With 899,840 LUTs and 1,968 DSP blocks available, a rough estimation can be derived in an ideal scenario \textbf{utilizing LUTs}, up to 449,920 additions and bit shifts could theoretically be implemented. This estimation assumes a simplistic model where each operation requires only 2 LUTs. However, it's important to note that this is an oversimplification, and actual resource utilization may vary due to practical considerations such as routing constraints. Additionally, \textbf{utilizing DSP blocks}, designed for efficient arithmetic operations, can potentially accommodate an additional 1,968 operations. However, this estimation presupposes optimal utilization of LUTs, which may not always be achievable in practice.  Furthermore, it's important to acknowledge that the actual feasibility of implementing NABs on FPGA hardware is influenced by factors such as clock speed, memory requirements, and specific application demands. Therefore, while these estimations provide valuable insights, they serve as initial benchmarks and must be validated through comprehensive analysis and optimization efforts tailored to the specific hardware platform and application requirements.

Finally, this work especially focuses on the four inference complexity metrics of the NN (RM, BOP, NABS, and NLG). Apart from the aforementioned metrics, the power consumption, memory footprint, and complexity of the activation function should be assessed. Power consumption should be considered as it can result in a bottleneck during the implementation phase. Memory footprint refers to the input size of the time series and the number of parameters that need to be saved in the memory, taking into account the quantization scheme. This complexity of the activation function, considering different types of activation functions and the approximation techniques, is another aspect to keep in mind  \cite{freire2023implementing}. 

\subsection{Usage of the complexity metrics in photonic hardware}
It can be questionable if the aforementioned metrics proposed to access the computational complexity in the inference phase of the electronic hardware can be used to access the complexity in the photonic hardware too. When discussing the complexity of computations in photonic NN hardware \cite{salmani2021photonic,zhou2022photonic} or photonics integrated circuits (e.g. PICs), it involves different considerations compared to traditional electronic computing (e.g., FPGA and ASIC). We can discuss metric by metric as follows: 
\begin{enumerate}
    \item Real multiplication is also suitable for photonic computing as in the context of NN, matrix-vector multiplications are a core operation. Photonic computing allows multiplications without the need for converting to electronic signals\cite{zhou2022photonic}.
    \item Number of bit operations (BOP) are only partially suitable for photonic computing because photonic computing can use the analog properties and continuous values of light, such as amplitude and phase, to encode information. However, digital photonic computing also exists. Digital photonic computing represents information in optical on-off keying or other discrete states.  BOP can be relevant but might not be the most comprehensive metric, depending on the types of photonic computing architectures.
    \item Number of adders and bit shifts (NABS) might not be a suitable metric because the photonic hardware does not directly use shifts and adders in the same way as electronic circuits. The operations in photonic systems are more about manipulating light's physical properties through optical components, for example, waveguides, modulators, and detectors.
    \item Number of logic gates is not a suitable metric for the photonic hardware, however, a similar concept can be considered. Instead of the gate count as in the electronic domain,  we can count the number of optical components like modulators, detectors, and waveguides in the photonic implementation. 
\end{enumerate}

It is important to consider other factors when assessing the trade-off between the performance and complexity of equalizers and the associated hardware. For instance, aspects such as power efficiency\cite{salmani2021photonic} and the comparison of physical size or chip area are crucial. Note, that a direct comparison of raw power consumption between PICs and ASICs might be misleading. To accurately assess power efficiency, we should consider throughput, which is the amount of data processed per unit of time. A more insightful metric is energy efficiency, obtained by dividing the power consumption by the achieved throughput. This allows for a fair comparison, highlighting the ability of each technology to perform a specific task with minimal energy consumption.

The chip area is another critical factor in many applications. When comparing the area occupied by a PIC and an ASIC equalizer, it is essential to consider the functionality delivered per unit area. A larger PIC might integrate functionalities beyond simple equalization, offering a higher overall value proposition despite its larger footprint\cite{bogaerts2020programmable}. Conversely, a smaller ASIC might excel in scenarios where compactness is paramount, even with a more limited functional range.

Finally, factors like scalability, reconfigurability, and sensitivity to fabrication imperfections can influence the choice between PICs and ASICs \cite{tan2022circuit}.

\section{Mathematical Complexity Formulation}\label{Sec:complexity}
In this section, we provide a brief introduction to various types of NN: dense layer, Convolutional Neural Networks (CNN),  Vanilla Recurrent Neural Networks (RNN), Long Short-Term Memory Neural Networks (LSTM), Gated Recurrent Units (GRU), and Echo State Networks (ESN). We investigate the computational complexity of each network in terms of RM, BOP, and NABS.
In this work, the computational complexity is formulated per layer, and the output layer is not taken into account for the complexity calculation to eliminate redundant computations if multiple layers or multiple NN types are combined. Table \ref{Table:Main} in the section's end summarizes the formulas for the RM, BOP, and NABS for all NN types studied. \textit{Furthermore, we have included all the complexity equations utilized in this study in Python code \cite{pedro_2024_10802496}. This resource is intended to aid readers in calculating the complexity of their own NN architectures.}

\begin{figure*}[ht]
\centering
\includegraphics[width=0.8\textwidth ]{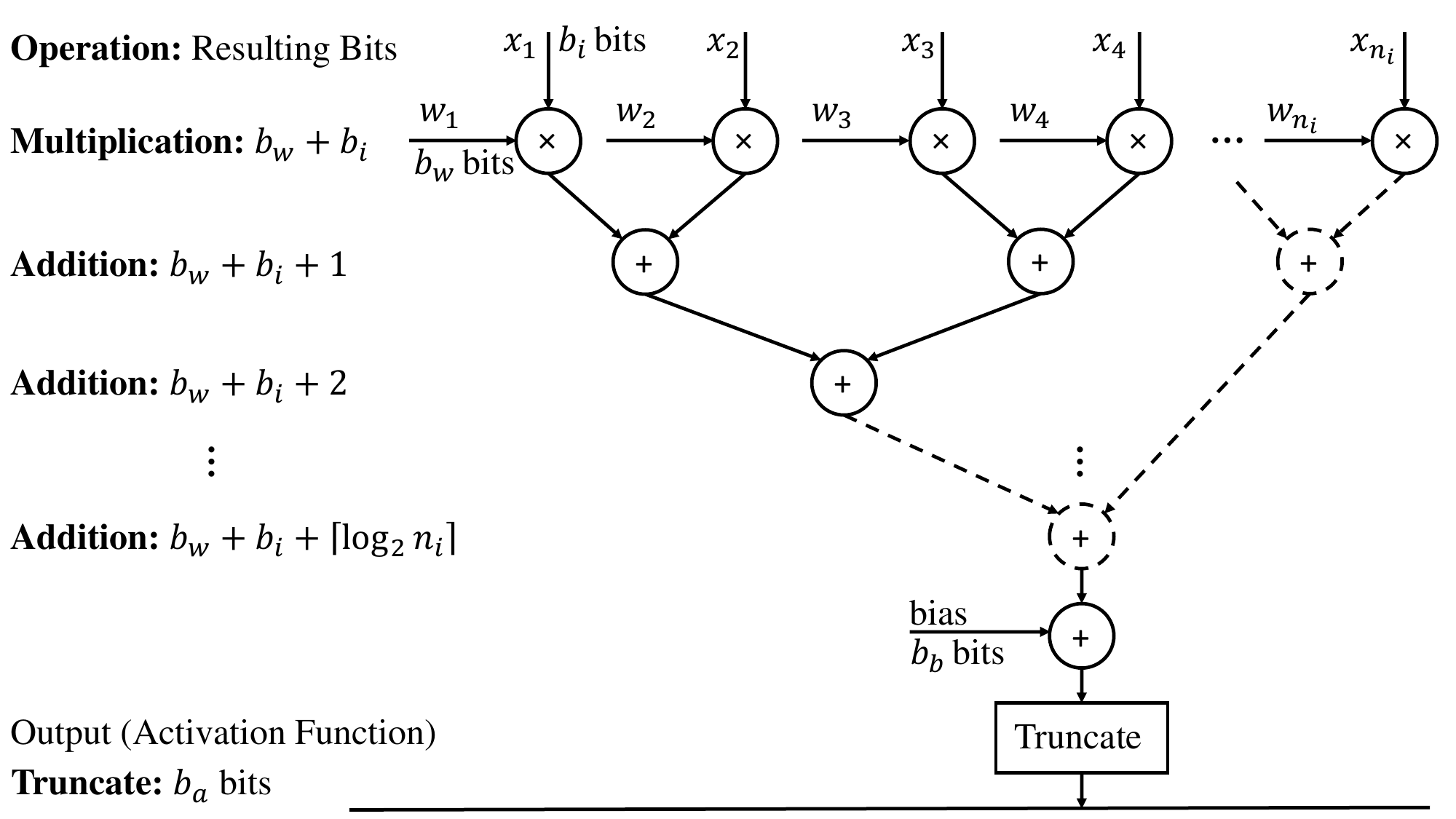}
\caption{Data path of a neuron in a quantized dense layer where $x$ is the input vector with size $n_i$, $w$ is the weight matrix, $b_w$ is the weight bitwidth and $b_i$ is the input bitwidth, $b_a$ is the activation bitwidth and $b_b$ is the bias bitwidth.}
\label{fig:dense_quantization}
\end{figure*}

\subsection{Dense Layer} \label{sec.dense}
A dense layer, also known as a `fully connected layer', is a layer in which each neuron is connected with all the neurons from the previous layer with a specific weight $w_{{i}{j}}$. The input vector is mapped to the output vector in a nonlinear manner by the dense layer, due to the participation of a non-linear activation function. Dense layers can be combined to form a Multi-Layer Perceptron (MLP), which is a class of a feed-forward deep NN. 

The output vector $y$ of a dense layer given $x$ as an input vector is written as:
\begin{equation}\label{eq.dense}
 y =\phi (Wx+ b ),
\end{equation}
where $y$ is the output vector, $\phi$ is a nonlinear activation function, $W$ is the weight matrix, and $b$ is the bias vector.
Writing explicitly the matrix operation inside the activation function:
  \begin{equation}\label{matrix.mul}
  \begin{split}
           W  x+b  =   
\begin{bmatrix}
           w_{11} & w_{12} & \hdots  & w_{1n_i}\\
           w_{21} & w_{22} & \hdots  & w_{2n_i}\\
           \vdots & \vdots& \hdots  & \vdots\\
           w_{n_n1} & w_{n_n2} & \hdots  & w_{{n_n}{n_i}}\\
         \end{bmatrix}
\begin{bmatrix} 
           x_{1}\\
           \! x_{2}\\
           \vdots\\
           x_{n_i}
         \end{bmatrix}\; 
        + \; \begin{bmatrix} 
           b_{1} \\
           b_{2} \\
           \vdots \\
           b_{n_{n}}
         \end{bmatrix} \!  ,
  \end{split}
  \end{equation}
where $n_i$ is the number of features in the input vector and $n_n$ represents the number of neurons in the layer, we can readily see that the RM of a dense layer can be computed according to the simple, well-known formula:
\begin{equation} \label{c.dense}
        \text{RM}_{\text{Dense}} = n_{n}n_{i}.
\end{equation}

Now we calculate the BOP of a dense layer, taking into account the bitwidth of two operands, to approximate the computational complexity of NNs when the mixed-precision arithmetic is used. The bitwidth, also known as the precision, is the number of bits used to represent a certain element; for example, each weight in the weight matrix can be represented with $b_w$ bits. Fig.~\ref{fig:dense_quantization} illustrates the data flow of the MAC operations for a neuron of a dense layer with $n_i$ input features and $b_i$ as input bitwidth. The multiplication of the input vector and the weights for one neuron can be mathematically represented as follows:

\begin{equation}
    y_{\text{MUL, one neuron}} = \sum_{n=1}^{n_i} w_n x_n. 
\end{equation}

Initially, the $n_i$ multiplications of input vector elements and weights for one neuron take place. When the multiplication of two operands is performed, the resulting bitwidth is the sum of the bitwidths of two operands ($b_w+b_i$) as shown in the first row of Fig.~\ref{fig:dense_quantization}.

After that, $n_i-1$ additions need to be made, and the resulting number of bits can be defined as follows: Considering that the result of the addition of two operands has the bitwidth of the bigger operand plus one bit, we start adding the multiplication results pairwise until only one element remains. In this case, the second row of Fig.~\ref{fig:dense_quantization} shows the first level of pairwise additions, with a resulting bitwidth of $b_w+b_i+1$, and since this pairwise addition process is repeated for $\lceil \log_2(n_{i})\rceil$ levels (until we have just a final single number), the total bitwidth of it is given by $b_{w} \!+ \!b_{i} +\lceil \log_2(n_{i}) \rceil$, i.e. it is the bitwidth required to perform the overall MAC process. Then, the addition of the bias vector is performed. In this work, for all types of networks, we assume that the size of the accumulator defined by the multiplication of the weight matrix and the input vector, is dominant; thereby, the assumption for the bias bitwidth $b_b$ is as follows: $b_b < b_{w} + b_{i} +\lceil \log_2(n_{i})\rceil$, and the addition of bias, in the end, will not result in the overflow. Finally, the bitwidth of the resulting number is truncated to $b_a$, where $b_a$ is the bitwidth of the activation function \cite{gysel2016hardware}.

When calculating the BOP for a dense layer, the costs of both multiplications and additions need to be included. Then, the BOP formula takes the form of the sum of two constituents, $\mathrm{BOP}_{\text{Mul}}$ and $\mathrm{BOP}_{\text{Bias}}$, corresponding to vector-matrix multiplication and bias addition:
\begin{equation} \label{BOP.dense.mul}
        \mathrm{BOP}_{\text{Mul}} = n_{n}\big[n_{i} b_{w} b_{i} + (n_{i}-1)(b_{w} + b_{i} +\lceil \log_2(n_{i}) \rceil)\big],
\end{equation}
\begin{equation} \label{BOP.dense.add}
        \mathrm{BOP}_{\text{Bias}} \approx n_{n}(b_{w} + b_{i} +\lceil \log_2(n_{i}) \rceil).
\end{equation}

Eq.~(\ref{BOP.dense.mul}) shows the cost of the number of one-bit full adders calculated from the dot product of $n_i$-dimensional input vector and weight matrix, as in Refs.~\cite{baskin2021uniq, tran2021ps}. The cost takes into account the bitwidths of the weights and input, $b_w$ and $b_i$. To compute the product of the two operands, we have to use $n_i n_n$ multiplications and $n_n (n_i-1)$ additions. The multiplication cost can be calculated by the number of multiplications multiplied by $b_w b_i$, which is related to the bit operation, and the number of additions multiplied by the accumulator bitwidth required to do the operation. The final BOP is the contribution of multiplication and the addition of bias of the dense layer. For the convenience of the forthcoming presentation, let us define the short notations: 
\[ \text{Mult}(n_{i}, b_{w}, b_{i}) = n_{i} b_{w} b_{i} + (n_{i}\!-\!1)\big(b_{w} \!+ \!b_{i} \!+\lceil \log_2(n_{i}) \!\rceil \big),\] 
and 
\[ \text{Acc}(n_{i}, b_{w}, b_{i}) = b_{w} + b_{i} +\lceil \log_2(n_{i}) \rceil.\] 
The Acc expression represents the actual bitwidth of the accumulator required for MAC operation, as shown in Fig. \ref{fig:dense_quantization}. Then, the BOP of the dense layer expressed through the layer parameters becomes:
\begin{equation} \label{BOP.dense}
\begin{split}
        \mathrm{BOP}_{\text{Dense}}
={}& \text{BOP}_{\text{Mul}} + \text{BOP}_{\text{Bias}}\\
\approx{}& n_{n}n_{i}\big[b_{w} b_{i} + (b_{w} + b_{i} +\lceil \log_2(n_{i}) \rceil)\big]\\
\approx{}& n_{n}n_{i}\big[b_{w} b_{i} + \text{Acc}(n_{i}, b_{w}, b_{i})\big].
\end{split}
\end{equation}

 Now, we note that with the advancement in NN quantization techniques, there arises the opportunity to approximate multiplication by using shift and few add operations only while still maintaining a good processing accuracy, since the NNs can diminish the approximation error that the quantized approximation introduces\footnote{Note that using the shifts and adders to perform multiplications can cause some quantization noise/error since we are converting from a float-point representation to a fixed-point representation with some defined quantized level of values. However, in NNs, this noise can be partially mitigated by including those quantized weights in the NN training process as in Refs.~\cite{Koike2021,elhoushi2021deepshift, you2020shiftaddnet}}
 \cite{Padmajarani2015FPGAIO,evans1994efficient}. As mentioned in Sec.~\ref{Sec:metrics}, the number of shifts can be neglected compared to the contribution of adders. The number of adders is different for different types of quantization. To be more specific, let $X$ represent the number of adders required, at most, to perform the multiplication and let $b$ be the bitwidth of the quantized matrix. For uniform quantization, we have $X = b-1$. And, for example, when the weight matrix with bitwidth of $b_{w}$, is quantized, we have $X_{w} = b_{w} - 1$ as the number of adders we need at most to perform the multiplication of the weights\footnote{Note that we can consider other techniques for the representation of such fixed-point multiplication to reduce its complexity e.g. the double-base number system where each multiplication with $b$ bits, at worst, needs no more than $b/log(b)$ additions \cite{dimitrov2010area}. For the Canonical Signed  Digit (CSD) representation, in the worst-case scenario, we have $(b + l)/2$ nonzero bits and on average it tends asymptotically to $(3b + l)/9$ \cite{hartley1996subexpression,dimitrov2007multiplication}}. In the case of Power-of-Two (PoT) quantization, we have $X = 0$, because each multiplication costs just a shift \cite{gentili1995efficient, przewlocka2022power}. Lastly, for the Additive Powers-of-Two (APoT) quantization, we have $X = n$, where $n$ denotes the number of additive terms. In APoT, the sum of $n$ PoT terms is used to represent each quantization level \cite{li2019additive}. Eventually, the NABS of a dense layer can be derived from its BOP equation, Eq.~(\ref{BOP.dense}):
\begin{equation} \label{nabs.dense}
\begin{split}
    \mathrm{NABS_{Dense}} 
\approx{}& n_{n}n_{i}\big[X_{w} \text{Acc}(n_{i}, b_{w}, b_{i})\! +\! \text{Acc}(n_{i}, b_{w}, b_{i})\big]\\
\approx{}& n_{n}n_{i}(X_{w} +1)\text{Acc}(n_{i}, b_{w}, b_{i}).
\end{split}
\end{equation}
As in Eq.~(\ref{nabs.dense}), the multiplication term $b_{w} b_{i}$ in Eq.~(\ref{BOP.dense}) is converted into the number of adders needed to operate the multiplication times the accumulator bitwidth required: $X_w\text{Acc}(n_{i}, b_{w}, b_{i})$.

\subsection{Convolutional Neural Networks}
In CNN, we apply the convolutions with different filters to extract the features and convert them into a lower-dimensional feature set, while still preserving the original properties. CNNs can be used in 1D, 2D, or 3D networks, depending on the applications. In this paper, we focus on 1D-CNNs, which apply to processing sequential data \cite{kiranyaz20211d}. 
For simplicity of understanding, the 1D-CNN processing with padding equal to 0, dilation equal to 1, and stride equal to 1, can be summarized as follows:
\begin{equation}\label{eq.cnn}
  y^{f}_{i} = \phi \left(\sum_{n=1}^{n_i}\sum_{j=1}^{n_k}x^{in}_{i+j-1,n} \cdot k^{f}_{j,n} + b^{f} \right),
\end{equation}
where $y^{f}_{i}$ denotes the output, known as a feature map, of a convolutional layer built by the filter $f$ in the $i$-th input element, $n_k$ is the kernel size, $n_i$ is the size of the input vector, $x^{in}$ represents the raw input data, $k^{f}_{j}$ denotes the $j$-th trainable convolution kernel of the filter $f$ and $b^{f}$ is the bias of the filter $f$.

In the general case, when designing the CNN, parameters like padding, dilation, and stride also affect the output size of the CNN. It can be formulated as:
\begin{equation}\label{outputwidth.cnn}
\mathrm{OutputSize} = \! \left[ \frac{n_s +2  \, padding \! - \! dilation (n_k-1) \! - \! 1 }{stride} \! + \! 1\right] \! ,
\end{equation}
where $n_s$ is the input time sequence size.

The RM of a 1D-convolutional layer can be computed as follows:
\begin{equation}\label{c.cnn}
    \mathrm{RM}_{\text{CNN}}= n_{f} n_{i} n_{k} \cdot \mathrm{OutputSize},
\end{equation}
where $n_{f}$ is the number of filters, also known as the output dimension. As in Eq.~(\ref{c.cnn}), there are $n_{i}n_{k}$ multiplications per sliding window, and the number of times that sliding window process needs to be repeated is equal to the output size. Then, the procedure is executed repeatedly for all $n_f$ filters.
 
The BOP for a 1D-convolutional layer, after taking into consideration the multiplications and additions, can be represented as:
\begin{equation} \label{BOP.cnn}
\begin{split}
        \mathrm{BOP}_{\text{CNN}}
={}& \mathrm{OutputSize} \cdot n_{f}\text{Mult}(n_{i}n_{k}, b_{w}, b_{i})\\
+{}& n_{f}\text{Acc}(n_{i}n_{k}, b_{w}, b_{i}).
\end{split}
\end{equation}
Eq.~(\ref{BOP.cnn}) is derived from Eq.~(\ref{eq.cnn}) and Eq.~(\ref{c.cnn}). The first term is associated with the convolution operation between the flattened input vector and the sliding windows, and the latter term corresponds to the addition of the bias.

The procedure to derive the NABS is similar to that described in detail in the case of a dense layer, Sec.~\ref{sec.dense}. The NABS of a 1D-convolutional layer is given by:
\begin{equation} \label{nabs.cnn}
\begin{split}
    \mathrm{NABS_{CNN}} 
={}& \mathrm{OutputSize} \cdot n_{f}\big[n_{i}n_{k}(X_{w} + 1 )-1\big]\\ 
{}&\cdot\text{Acc}(n_{i}n_{k}, b_{w}, b_{i})\\
+{}& n_{f}\text{Acc}(n_{i}n_{k}, b_{w}, b_{i}).\\
\end{split}
\end{equation}
To obtain the 1D-convolutional layer's NABS, the multiplication in Eq.~(\ref{BOP.cnn}) is represented by the number of adders required, at most, to perform the multiplication times the accumulator bitwidth.

\subsection{Vanilla Recurrent Neural Networks} \label{sec.rnn}
Vanilla RNN is different from MLP and CNN in terms of its ability to handle memory, which is quite beneficial for time series data. RNNs take into account the current input and the output that the network has learned from the prior input. Even though the RNNs introduced efficient memory handling, they still suffer from the inability to capture the long-term dependencies because of the vanishing gradient issue \cite{bengio1994vanishgrad}. 
The equation for the vanilla RNN given a time step $t$ is as follows:
\begin{equation}\label{eq.rnn}
    \begin{gathered}
    h_{t} = \phi(W{x}_{t} + Uh_{t-1} + b), \\
    \end{gathered}
\end{equation}
where $\phi$ is, again, the nonlinear activation functions, $x_{t}\in \mathbb{R}^{n_i}$ is the $n_i$-dimensional input vector at time $t$, $h_{t} \in \mathbb{R}^{n_h}$ is a hidden layer vector of the current state with size $n_h$,  $W \in \mathbb{R}^{n_h \times n_i}$  and $U\in \mathbb{R}^{n_h \times n_h}$ represent the trainable weight matrices, and $b$ is the bias vector. For more explanations on the vanilla RNN operation, see Ref.~\cite{lipton2015critical}.
The RM of a vanilla RNN is:
\begin{equation} \label{c.rnn}
        \mathrm{RM}_{\text{RNN}} = n_{s}n_{h}(n_{i} + n_{h}),
\end{equation}
where $n_h$ notes the number of hidden units. From Eq.~(\ref{c.rnn}), the RM for a time step is $n_{h}(n_{i} + n_{h})$.  It can be separated into two terms; the $n_{h}n_{i}$ term corresponds to the multiplication of the input vector $x_t$ and the weight matrix, and the $n_{h}^2$ term arises because of the multiplication to the prior cell output $h_{t-1}$. Finally, $n_{s}$, which denotes the number of time steps in the layer, should be taken into account, as the process is repeated $n_s$ times.

The BOP for a vanilla RNN is given as:
\begin{equation} \label{BOP.rnn}
\begin{split}
        \mathrm{BOP}_{\text{RNN}}
={}& n_{s}n_{h}\text{Mult}(n_{i}, b_{w}, b_{i})\\
+{}& n_{s}n_{h}\text{Mult}(n_{h}, b_{w}, b_{a})\\
+{}& 2n_{s}n_{h}\text{Acc}(n_{h}, b_{w}, b_{a}).
\end{split}
\end{equation}
From Eq.~(\ref{BOP.rnn}), the first term is associated with the input vector multiplied by the weight matrix, and the second term corresponds to the multiplications of the recurrent cell outputs. The final term is the contribution of the addition between $W{x}_{t} + Uh_{t-1}$ and the addition of the bias vector in Eq.~(\ref{eq.rnn}); one can see that the size of the accumulator used in this term, is $\text{Acc}(n_{h}, b_{w}, b_{a})$. It is due to the assumption that $\text{Acc}(n_{h}, b_{w}, b_{a})$ is dominant because it should be greater than $\text{Acc}(n_{i}, b_{w}, b_{i})$ as a result of the inequality $n_h > n_i$. \

As in the case of a dense layer, the NABS of vanilla RNN can be calculated from its BOP equation by converting the multiplication to the number of adders needed at most ($X$) depending on the quantization scheme and the accumulator size:
\begin{equation} \label{nabs.rnn}
\begin{split}
    \mathrm{NABS_{RNN}} 
={}& n_{s}n_{h}\big[ n_{i}(X_{w} + 1) -1\big]\text{Acc}(n_{i}, b_{w}, b_{i})\\
+{}& n_{s}n_{h}\big[ n_{h}(X_{w} + 1) +1\big]\text{Acc}(n_{h}, b_{w}, b_{a}).
\end{split}
\end{equation}

\subsection{Long Short-Term Memory Neural Networks} \label{sec.lstm}
LSTM is an advanced type of RNNs. Although RNNs suffer from short-term memory issues, the LSTM network can learn long-term dependencies between time steps ($t$), insofar as it was specifically designed to address the gradient issues encountered in RNNs \cite{hochreiter1997long, gers2000learning}. There are three types of gates in an LSTM cell: an input gate ($i_t$), a forget gate ($f_t$), and an output gate ($o_t$). More importantly, the cell state vector ($C_t$) was proposed as a long-term memory to aggregate the relevant information throughout the time steps.
The equations for the forward pass of the LSTM cell given a time step $t$ are as follows:
\begin{equation}\label{eq.lstm}
    \begin{gathered}
i_{t} = \sigma(W^{i}{x}_{t} + U^{i}{h}_{t-1} + b^{i} ),  \\
    f_{t} = \sigma(W^{f}{x}_{t} + U^{f}{h}_{t-1} + b^{f}), \\
o_{t} = \sigma(W^{o}{x}_{t} + U^{o}{h}_{t-1} + b^{o}),\\
    C_{t} = f_{t}\odot C_{t-1} + i_{t}\odot \phi(W^{c}{x}_{t} + U^{c}{h}_{t-1}+ b^{c}), \\
    h_{t} = o_{t} \odot \phi(C_{t}),
    \end{gathered}
\end{equation}
where $\phi$ is usually the ``tanh'' activation function, $\sigma$ is usually the sigmoid activation function, the sizes of each variable are  $x_{t}\in \mathbb{R}^{n_i}$,  $f_{t}, i_{t}, o_{t}\in (0,1)^{n_h}$, $C_{t}\in \mathbb{R}^{n_h}$ and $h_{t}\in (-1,1)^{n_h}$. The $\odot$ symbol represents the element-wise (Hadamard) multiplication. 

The RM of an LSTM layer is:
\begin{equation} \label{c.lstm}
    \mathrm{RM}_{\text{LSTM}} = n_{s}n_{h}(4n_{i} + 4n_{h} + 3),
\end{equation}
where $n_h$ is the number of hidden units in the LSTM cell. Similarly to RNNs, the RM can be calculated from the term associated with the input vector $x_t$ and the term corresponding to the prior cell output ${h}_{t-1}$; however, each term occurs four times, as we can see in Eq.~(\ref{eq.lstm}). Therefore, we have $4n_hn_i$ and $4n_h^2$, respectively. Moreover, we also need to include the element-wise product that is operated three times in Eq.~(\ref{eq.lstm}), which costs $3n_h$. Finally, the process is repeated $n_s$ times, hence, $n_s$ is multiplied to the overall number. 

The BOP for an LSTM layer is computed based on Eq.~(\ref{c.lstm}), but also includes the bitwidth of the operands and the number of additions. As a result, the BOP can be represented as:
\begin{equation} \label{BOP.lstm}
\begin{split}
        \mathrm{BOP}_{\text{LSTM}}
={}& 4n_{s}n_{h}\text{Mult}(n_{i}, b_{w}, b_{i})\\
+{}& 4n_{s}n_{h}\text{Mult}(n_{h}, b_{w}, b_{a})\\
+{}& 3n_{s}n_{h}b_{a}^2\\
+{}& 9n_{s}n_{h}\text{Acc}(n_{h}, b_{w}, b_{a}).
\end{split}
\end{equation}
 To give more details on the expression, the first two terms in Eq.~(\ref{BOP.lstm}) are the contribution of the input vector multiplications and the recurrent cell output association, respectively. The term $3n_{s}n_{h}b_{a}^2$ refers to 3 times of the element-wise product of two operands with $b_a$ bitwidth, see Eq.~(\ref{eq.lstm}). For each time step, there are $n_h$ elements in each vector that need to be multiplied. The last term is for all the additions since we assume that $\text{Acc}(n_{h}, b_{w}, b_{a})$ gives the dominant contribution, as described in Sec.~\ref{sec.rnn}. Finally, the process is then restarted $n_s$ times.

The NABS of an LSTM layer is derived from the Eq.~(\ref{BOP.lstm}) by replacing the multiplications with the shifts and adders including their cost, as mentioned in Sec.~\ref{Sec:metrics} that the shifts would not be included. The number of adders depends on the quantization technique. The NABS would be as follows:
\begin{equation} \label{nabs.lstm}
\begin{split}
    \mathrm{NABS_{LSTM}} 
={}& 4n_{s}n_{h}\big[n_{i}(X_{w} +1)-1\big]\text{Acc}(n_{i}, b_{w}, b_{i})\\
+{}& 4n_{s}n_{h}\big[n_{h}(X_{w} +1) +1\big]\text{Acc}(n_{h}, b_{w}, b_{a})\\
+{}& 6n_{s}n_{h}b_{a}.
\end{split}
\end{equation}
The first and second terms are the results of input vector multiplications and the recurrent cell operation combined with all addition operations, respectively. In this case, the third term comes from $3n_{s}n_{h}(b_{a}+b_{a})$. Due to the element-wise product of two operands with bitwidth $b_a$, the resulting bitwidth becomes $b_a+b_a$ as mentioned in Fig. \ref{fig:dense_quantization}.

\subsection{Gated Recurrent Units}
Like LSTM, the GRU network was created to overcome the short-term memory issues of RNNs. However, GRU is less complex, as it has only two types of gates: reset ($r_t$) and update ($z_t$) gates. The reset gate is used for short-term memory, whereas the update gate is responsible for long-term memory \cite{dey2017gate}. In addition, the candidate hidden state ($h'_{t}$) is also introduced to state how relevant the previous hidden state is to the candidate state.
The GRU for a time step $t$ can be formalized as: 
\begin{equation}\label{eq.gru}
    \begin{gathered}
    z_{t} = \sigma(W^{z}{x}_{t} + U^{z}{h}_{t-1} + b^{z}),  \\
    r_{t} = \sigma(W^{r}{x}_{t} + U^{r}{h}_{t-1} + b^{r}), \\
    h'_{t} = \phi(W^{h}{x}_{t} + r_{t} \odot U^{h}{h}_{t-1} + b^{h}), \\
    h_{t} = z_{t} \odot {h}_{t-1} + (1 - z_{t}) \odot h'_{t},
    \end{gathered}
\end{equation}
where $\phi$ is typically the ``tanh'' activation function and the rest of the designations are the same as in Eq.~(\ref{eq.lstm}).

The RM of the GRU is calculated in the same way as we did for the LSTM in Eq.~(\ref{c.lstm}), but the number of operations with the input vector $x_t$ and with the previous cell output ${h}_{t-1}$ is reduced from four (LSTM) to three times as shown in Eq.~(\ref{eq.gru}). Thus, the expression for the RM becomes:
\begin{equation} \label{c.gru}
        \mathrm{RM}_{\text{GRU}}= n_{s}n_{h}(3n_{i} + 3n_{h} + 3).
\end{equation}

The BOP for the GRU can be calculated in the same manner as we did for the LSTM in Eq.~(\ref{BOP.lstm}). However, now, the expression is slightly different in the number of matrix multiplications as the number of gates is now lower. The BOP number can be represented as:
\begin{equation} \label{BOP.gru}
\begin{split}
        \mathrm{BOP}_{\text{GRU}}
={}& 3n_{s}n_{h}\text{Mult}(n_{i}, b_{w}, b_{i})\\
+{}& 3n_{s}n_{h}\text{Mult}(n_{h}, b_{w}, b_{a})\\
+{}& 3n_{s}n_{h}b_{a}^2\\
+{}& 8n_{s}n_{h}\text{Acc}(n_{h}, b_{w}, b_{a}).
\end{split}
\end{equation}
The explanation for each line here is identical to that in Eq.~(\ref{BOP.lstm}).

The NABS of the GRU is derived similarly to the LSTM case:
\begin{equation} \label{nabs.gru}
\begin{split}
    \mathrm{NABS_{GRU}} 
={}& 3n_{s}n_{h}\big[n_{i}(X_{w} + 1)-1\big]\text{Acc}(n_{i}, b_{w}, b_{i})\\
+{}& n_{s}n_{h}\big[3n_{h}(X_{w} + 1)+5\big]\text{Acc}(n_{h}, b_{w}, b_{a})\\
+{}& 6n_{s}n_{h}b_{a}.
\end{split}
\end{equation}
Again, the explanation for each term in this expression is identical to Eq.~(\ref{nabs.lstm}).

\subsection{Echo State Networks} \label{sec.esn}
ESN belongs to the class of recurrent layers, but more specifically, to the reservoir computing category. ESN was proposed to relax the training process while being efficient and simple to implement. The ESN comprises three layers: an input layer, a recurrent layer, known as a reservoir, and an output layer, which is the only layer that is trainable. The reservoir with random weight assignment is used to replace back-propagation in traditional NNs to reduce the computational complexity of training \cite{wu2018statistical}. We notice that the reservoir of the ESNs can be implemented in two domains: digital and optical \cite{sorokina2019fiber}. With the optical implementation of the reservoir, the computational complexity dramatically falls, however, the degradation of the performance due to the change of domain is noticeable \cite{brain-inspired}. In this work, we only examine the digital domain implementation. Moreover, we focus on the leaky-ESN, as it is believed to often outperform standard ESNs and is more flexible due to time-scale phenomena \cite{sun2020review,jaeger2007optimization}. 
The equations of the leaky-ESN for a certain time step $t$ are given as:
\begin{equation}\label{eq.esn1}
 a_t = \phi \left( W^{r} s_{t-1} + W^{\text{in}} x_t \right),
\end{equation}
\begin{equation}\label{eq.esn2}
  s_t = (1- \mu) s_{t-1} + \mu a_t,
\end{equation}
\begin{equation}\label{eq.esn3}
y_t = W^{o}s_{t} + b^{o},
\end{equation}
where $s_t$ represents the state of the reservoir at time $t$, $W^r$ denotes the weight of the reservoir with the sparsity parameter $s_p$, $W^{in}$ is the weight matrix that shows the connection between the input layer and the hidden layer, $\mu$ is the leaky rate, $W^{o}$ denotes the trained output weight matrix, and $y_t$ is the output vector.

The RM of an ESN is given by
\begin{equation}\label{c.esn}
    \mathrm{RM}_{\text{ESN}}= n_sN_{r}(n_{i} + N_{r}s_{p}+2 + n_{o}),
\end{equation}
where $N_r$ is the number of internal hidden neuron units of the reservoir and $n_o$ denotes the number of output neurons. From Eq.~(\ref{c.esn}), the $N_rn_i$ multiplications occur from the input vector operations, and the term $N_r^{2}s_p$ is included due to the reservoir layer; to be more specific, the latter term is multiplied with the sparsity parameter $s_p$ which indicates the ratio of zero values in the matrix. Eq.~(\ref{eq.esn2}) results in $2N_r$ multiplications. Unlike the other network types, now we have to include the contribution of the output layer explicitly because it contains the trainable weight matrix, and this layer contributes $N_r n_o$ multiplications. Eventually, the process is repeated for $n_s$ times.

The BOP number for an ESN can be represented as:
\begin{equation} \label{BOP.esn}
\begin{split}
        \mathrm{BOP}_{\text{ESN}}
={}& n_{s}N_{r}\text{Mult}(n_{i}, b_{w}, b_{i})\\
+{}& n_{s}N_{r}s_{p}\text{Mult}(N_{r}, b_{w}, b_{a})\\
+{}& n_{s}N_{r}\text{Mult}(n_{o}, b_{w}, b_{a})\\
+{}& 2n_{s}N_{r}b_{a}^2\\
+{}& 4n_{s}N_{r}\text{Acc}(N_{r}, b_{w}, b_{a}).
\end{split}
\end{equation}
In Eq.~(\ref{BOP.esn}), the first term is the input vector contribution, the second one is contributed by the reservoir layer, the third term refers to the output layer multiplications, and the fourth term stems from the multiplications in Eq.~(\ref{eq.esn2}). Eventually, all the addition operations are accounted for by the final term.

The NABS of an ESN, which can be calculated similarly as in the LSTM case in Sec.~\ref{sec.lstm}, is:
\begin{equation} \label{nabs.esn}
\begin{split}
    \mathrm{NABS_{ESN}} 
={}& n_{s}N_{r}\big[n_{i}(X_{w} + 1) -1\big]\text{Acc}(n_{i}, b_{w}, b_{i})\\
+{}& n_{s}N_{r}\big[s_{p}(N_{r}X_{w} + N_{r} -1\big] +4 )\text{Acc}(N_{r}, b_{w}, b_{a})\\
+{}& n_{s}N_{r}\big[n_{o}(X_{w} + 1) -1\big]\text{Acc}(n_{o}, b_{w}, b_{a})\\
+{}& 4n_{s}N_{r}b_{a}.
\end{split}
\end{equation}
By changing the multiplication terms in Eq.~(\ref{BOP.esn}) to the number of adders required at most, we obtain the ESN's NABS. The input vector multiplication contributes to the first term. The reservoir layer and all the addition operations result in the second term. The third term comes from the output layer of the ESN. The last term is the contribution of Eq.~(\ref{eq.esn2}).

\subsection{Residual Neural Network} \label{sec.resnet}
Residual Neural Network (ResNet)\cite{he2016deep} is a feed-forward NN architecture with shortcut connections to skip one or more layers and perform the identity mapping. ResNet addresses vanishing gradient and the degradation problems\cite{veit2016residual, he2016deep}. The degradation problems relate to the saturation and the rapid degradation of the accuracy of the deeper networks when they start to converge. 
It has shown remarkable performance in computer vision tasks. ResNet primarily utilizes 2D-convolutional layers as its core building blocks to extract features. The output of ResNet building block, see Fig.~\ref{fig:resnet}, is derived as follows:
\begin{figure}[t]
    \centering
    \includegraphics[scale=0.4]{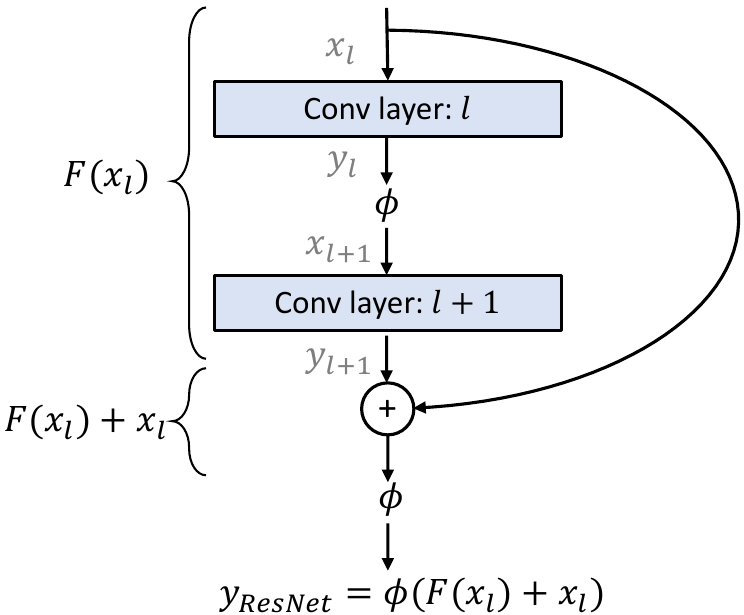}
    \caption{Architecture of a ResNet block.}
    \label{fig:resnet}
\end{figure}
\begin{equation} \label{eq.resnet}
\begin{split}
    y_\mathrm{ResNet} 
={}& \phi(F(x_l)+x_l)\\
+{}& \phi(x_{l+1}*k_{l+1}+x_l)\\
+{}& \phi(\phi(y_l)*k_{l+1}+x_l)\\
+{}& \phi(\phi(x_l*k_l)*k_{l+1}+x_l),
\end{split}
\end{equation}
where $\phi$ is the activation function which, in the original paper, is the linear rectifier unit (ReLU) 

For simplicity of the calculation for each layer, the 2D-convolution output without the activation function, assumed padding equal to 0, dilation equal to 1, and stride equal to 1, can be represented as follows:
\begin{equation}\label{eq.cnn2d}
  y^{f}_{i,j} = (x^{in}*k^f)_{i,j} = \sum_{p=1}^{n_{i}}\sum_{m=1}^{n_{k_1}}\sum_{n=1}^{n_{k_2}}x^{in}_{i+m,j+n,p} \cdot k^{f}_{m,n,p} + b^{f},
\end{equation}
where $y^{f}_{i,j}$ denotes the output element of the feature map, of a convolutional layer built by the filter $f$, with $i$ and $j$ representing the row and column indices of the output element, 
$x^{in} \in \mathbb{R}^{n_i \times n_s}$ represents the raw input data and $n_{k_1}$ and $n_{k_2}$ are the dimension of the 2D kernel $k^f$.

The size of the feature map or convolution output, in fact, depends on other parameters like padding, dilation, and stride. The width and height of the feature map are formulated as follows: 
\begin{equation}\label{outputwidth.cnn2d}
    \begin{gathered}
    \mathrm{Width} = \! \left[ \frac{n_{x} +2  \, padding \! - \! dilation (n_{k_1}-1) \! - \! 1 }{stride} \! + \! 1\right] \!,  \\
    \mathrm{Height} = \! \left[ \frac{n_s +2  \, padding \! - \! dilation (n_{k_2}-1) \! - \! 1 }{stride} \! + \! 1\right] \!, \\
    \mathrm{OutputSize}=\mathrm{Width}\times \mathrm{Height},
    \end{gathered}
\end{equation}
where $n_s$ is the first dimension of the input and $n_x$ is the second dimension of the input. In the case of 2D CNN, the $n_s$ does not refer to the time step because 2D CNN is usually applied to the image processing rather than the time series data. Note that $x^{in} \in \mathbb{R}^{n_x \times n_s \times n_i}$ Therefore, we have $y^f\in \mathbb{R}^{\mathrm{Width} \times \mathrm{Height}}$.

The RM of each 2D-convolutional layer can be calculated by
\begin{equation}\label{c.2dcnn}
    \mathrm{RM}_{\text{2D-CNN}}= n_{f} n_{i} n_{k_1} n_{k_2} \cdot \mathrm{OutputSize},
\end{equation}
where $n_{f}$ is the number of filters (the output dimension) and $n_{i}$ is the number of features. For each feature, Eq.~(\ref{eq.cnn2d}), consists of $n_{k_1}n_{k_2}$ multiplications per sliding window, and the number of times the sliding window process needs to be repeated is equal to the output size. The calculation is repeated for each feature, in total $n_{i}$ times. Then, the whole procedure is repeated for all $n_f$ filters.

The ResNet block consists of several convolutional layers, in this case, two layers. Eq.~(\ref{eq.resnet}) shows two convolution operations. The RM of a ResNet block is given by the summation of $\mathrm{RM}_{\text{2D-CNN}}$ of each layer as follows:
\begin{equation}\label{c.resnet}
    \mathrm{RM}_{\text{ResNet}}= \mathrm{RM}^{l}_{\text{2D-CNN}}+\mathrm{RM}^{l+1}_{\text{2D-CNN}},
\end{equation}
where $l$ and $l+1$ represent the first and second layers in the ResNet block, respectively. 

The BOP of each 2D-convolutional layer can be formulated as:
\begin{equation} \label{BOP.2dcnn}
\begin{split}
        \mathrm{BOP}_{\text{2D-CNN}}
={}& \mathrm{OutputSize} \cdot n_{f}\text{Mult}(n_{i}n_{k_1}n_{k_2}, b_{w}, b_{i})\\
+{}& n_{f}\text{Acc}(n_{i}n_{k_1}n_{k_2}, b_{w}, b_{i}).
\end{split}
\end{equation}
The first term refers to the multiplication of the kernel and the input, and the second term refers to the bias addition. Please note that $b_i$ is the bitwidth of the layer input. When connecting the layer in sequence, the bitwidth of the layer can be the result of the previous layer, or usually it is quantized as a result of the activation function. 

The BOP of the ResNet block can be calculated from the BOP of two 2D-convolutional layers and the addition of the input (skip connection) as: 
\begin{equation}\label{bop.resnet}
\begin{split}
        \mathrm{BOP}_{\text{ResNet}}={}&\mathrm{BOP}^{l}_{\text{2D-CNN}}+\mathrm{BOP}^{l+1}_{\text{2D-CNN}}\\
        +{}&\text{Acc}(n_{i}n_{k_1}n_{k_2}, b_{w}, b_{i}).
\end{split}
\end{equation}
The addition term of the skip connection has BOP complexity of the larger operand which is the bitwidth of the convolution of the layer $l+1$, as can be observed in Eq.~(\ref{eq.resnet}). 

The NABS of the 2D-convolutional layer is formulated as follows:
\begin{equation} \label{nabs.2dcnn}
\begin{split}
\mathrm{NABS}_{\text{2D-CNN}}
= {}&\mathrm{OutputSize} \cdot n_{f}\left[n_{i}n_{k_1}n_{k_2}(X_w+1)-1\right]\\
\cdot{}& \text{Acc}(n_{i}n_{k_1}n_{k_2}, b_{w}, b_{i}) \\
+{}&n_{f}\text{Acc}(n_{i}n_{k_1}n_{k_2}, b_{w}, b_{i}).
\end{split}
\end{equation}
The NABS of a ResNet block is a combination of two 2D-convolutional layers and the addition of the skip connection, the same way as we calculate BOP:
\begin{equation}\label{nabs.resnet}
\begin{split}
        \mathrm{NABS}_{\text{ResNet}}={}&\mathrm{NABS}^{l}_{\text{2D-CNN}}+\mathrm{NABS}^{l+1}_{\text{2D-CNN}}\\
        +{}&\text{Acc}(n_{i}n_{k_1}n_{k_2}, b_{w}, b_{i}).
\end{split}
\end{equation}


\subsection{Transformer} \label{sec.transformer}
The Transformer \cite{vaswani2017attention} is a deep learning architecture where, instead of using recurrence, it leverages attention mechanisms to capture global long-range dependencies between input and output. The Transformer model has played a dominant role in natural language processing, and become a state-of-the-art model in various language-related tasks. The Transformer architecture comprises different layers; multi-head self-attention, masked multi-head attention, feed-forward, and Add\&Norm layer (residual connection and layer normalization). 

The attention function is computed from a matrix $Q$ which is a set of queries simultaneously, the matrix $K$ and $V$ for the keys and values. The dot products of the query with all keys are calculated and result in the output matrix:
\begin{equation}\label{eq.transformer}
    \mathrm{Attention}(Q,K,V) = \mathrm{softmax}(\frac{QK^T}{\sqrt{d_k}})V,
\end{equation}
where $\frac{1}{\sqrt{d_k}}$ is a scaling factor computed from keys of dimension $d_k$. The dimensions of each matrix are $Q \in \mathbb{R}^{m \times d_k}$, $K \in \mathbb{R}^{n \times d_k}$ and $V \in \mathbb{R}^{n \times d_v}$

For multi-head attention, it can be defined as:
\begin{equation}\label{eq.multihead}
\begin{gathered}
    \mathrm{MultiHead}(Q,K,V) = \mathrm{Concat(head_1,...,head_h)}W^O,\\
  \text{where} \;\mathrm{head_i} = \mathrm{Attention}(QW^Q_i,KW^K_i,VW^V_i),
\end{gathered}
\end{equation}
where $h$ is the number of times we calculate self-attention with different sets of $Q$, $K$, and $V$. To reduce the dimension of each head, the queries, keys, and values with different learned linear projections are projected to $d_k$, $d_k$, and $d_v$ dimensions, respectively. $d_v$ represents the dimensional output values. $W^Q_i \in \mathbb{R}^{d_{\text{model}} \times d_k}$, $W^K_i \in \mathbb{R}^{d_{\text{model}} \times d_k}$, $W^V_i \in \mathbb{R}^{d_{\text{model}} \times d_v}$, and $W^O \in \mathbb{R}^{hd_v \times d_{\text{model}}}$ are learnable weight matrices. 

In the masked multi-head attention, one way to mask out the inputs of the softmax function is to add the matrix $M$ which contains 0's and $-\infty$'s. The $-\infty$'s correspond to invalid connections. The equation then is modified to
\begin{equation}\label{eq.maskedattention}
   \mathrm{Attention_{Masked}}(Q,K,V) = \mathrm{softmax}(\frac{QK^T}{\sqrt{d_k}}+M)V
\end{equation}

The fully-connected feed-forward network (FFN) is to perform two linear transformations with a ReLU in between, as follows:
\begin{equation}\label{eq.ffn}
   \mathrm{FFN}(x) = \mathrm{ReLU}(xW_1+b_1)W_2+b_2.
\end{equation}
At the different positions where the linear transformations are applied, they use different parameters from layer to layer.  

Regarding the Add\&Norm layer, it involves a residual connection around each of the two sub-layers \cite{he2016deep} followed by layer normalization \cite{ba2016layer}. The output of this Add\&Norm layer is:
\begin{equation}\label{eq.addnorm}
   y_\text{{Add\&Norm}} = \mathrm{LayerNorm}(x+\text{Sublayer}(x)),
\end{equation}
where $\text{Sublayer}(x))$ refers to the function implemented by the sub-layer itself, for instance, FFN or MultiHead. The LayerNorm are then  defined as:
\begin{equation}\label{c.layernorm}
   \mathrm{LayerNorm}(z) = \frac{g(z-\mu_z)}{\sigma_z}+b,
\end{equation}
where $g$ and $b$ denote gain and bias, respectively, $\mu$ and $\sigma$ are the mean and variance of the summed inputs within each layer, respectively.

In this paper, we aim to calculate the RM, BOP, and NABS for each layer of the Transformer separately, to facilitate the custom-built transformer architecture. The RM of the multi-head attention equals to:
\begin{equation}\label{c.attention}
\begin{split}
   \mathrm{RM}_{\text{Multi-head}} = {}&h[ md_{\text{model}}d_k + nd_{\text{model}}d_k + nd_{\text{model}}d_v \\
   +{}&2mnd_{\text{model}}]+md_{\text{model}}hd_v,
\end{split}
\end{equation}
where $md_{\text{model}}d_k$, $nd_{\text{model}}d_k$ and $nd_{\text{model}}d_v $ result from multiplications of $QW^Q_i$
$KW^K_i$, and $VW^V_i$, respectively. The term $2mnd_{\text{model}}$ comes from the multiplications in Attention function. The last term results from the multiplication in the MultiHead function. 

The BOP of a multi-head layer is
\begin{equation}\label{bop.attention}
\begin{split}
   \mathrm{BOP}_{\text{Multi-head}} = {}&h\Big[ md_{\text{model}}\text{Mult}(d_k,b_Q,b_w)\\
   +{}&nd_{\text{model}}\text{Mult}(d_k,b_K,b_w)\\
   +{}&nd_{\text{model}}\text{Mult}(d_v,b_V,b_w)\\
   +{}&nm\text{Mult}(d_\text{model},b_{q},b_{q})\\
   +{}&md_{\text{model}}\text{Mult}(n,b_{a},b_{\text{q}})\Big]\\
   +{}&hd_vm\text{Mult}(d_{\text{model}},b_{q},b_w),
\end{split}
\end{equation}
where $b_Q$, $b_K$, $b_V$ and $b_w$ are the bitwidth of matrices Q, K, V, and W respectively. $b_{a}$ is the bitwidth of the activation function, in this case, softmax.  We assume that the resulting bitwidth from the matrix multiplications is quantized to $b_{\text{q}}$. The first, second, and third terms correspond to multiplications of $QW^Q_i$, $KW^K_i$, and $VW^V_i$, respectively. The fourth and fifth terms are the contributions of the Attention function. The last term results from the Multi-Head function. 
The NABS of a multi-head layer can be calculated as:
\begin{equation}\label{nabs.attention}
\begin{split}
   \mathrm{NABS}_{\text{Multi-head}} = {}&h\Big[ md_{\text{model}}\big[d_k(X+\!1)\!-\!1\big]\text{Acc}(d_k,b_Q,b_w)\\
   +{}&nd_{\text{model}}\big[d_k(X+1)-1\big]\text{Acc}(d_k,b_K,b_w)\\
   +{}&nd_{\text{model}}\big[d_v(X+1)-1\big]\text{Acc}(d_v,b_V,b_w)\\
   +{}&nm\big[d_\text{model}(X+1)-1\big]\text{Acc}(d_\text{model},b_{q},b_{q})\\
   +{}&md_{\text{model}}\big[n(X\!+1\!)\!\!-\!1\big]\text{Acc}(n,b_{a},b_{\text{q}})\Big]\\
   +{}&hd_vm\big[d_\text{model}(X+1)-1\big]\text{Acc}(d_{\text{model}},b_{q},b_w).
\end{split}
\end{equation}

For the point-wise FFN, The calculation is nearly the same as the dense layer. The RM can be formulated as:
\begin{equation}\label{c.ffn}
   \mathrm{RM}_{\text{FFN}} = 2d_{\text{model}}d_\text{ff}.
\end{equation}
It results from two times of the multiplication with weight matrices $W_1$ and $W_2$ in FFN. Each weight matrix for FNN has the size of $W_1 \in \mathbb{R}^{d_{\text{model}} \times d_\text{ff}}$ and $W_2 \in \mathbb{R}^{d_\text{ff} \times d_{\text{model}}}$. Note that the input vector has a dimension of $d_{\text{model}}$.

The BOP of the point-wise FFN is 
\begin{equation}\label{bop.ffn}
\begin{split}
\mathrm{BOP}_{\text{FNN}} = {}&d_{\text{model}}d_\text{ff} \Big[b_w b_i + \text{Acc}(d_{\text{model}}, b_w, b_i)\\
   +{}&  b_a b_i+ \text{Acc}(d_{\text{ff}}, b_a, b_i)\Big],
\end{split}
\end{equation}
where we assume that after the Relu activation function, the resulting bitwidth becomes $b_a$.

The NABS of the point-wise FFN can be found by:
 \begin{equation}\label{nabs.ffn}
\begin{split}
\mathrm{NABS}_{\text{FNN}} = {}&d_{\text{model}}d_\text{ff} \Big[(X_w+1)\text{Acc}(d_{\text{model}}, b_w, b_i)\\
   +{}&  (X_w+1)\text{Acc}(d_{\text{ff}}, b_a, b_w)\Big].
\end{split}
\end{equation}

Regarding the Add\&Norm layer, the calculation of the RM is relatively simple:
\begin{equation}\label{c.addnorm}
   \mathrm{RM}_{\text{Add\&Norm}} = n_x,
\end{equation}
where $n_x$ refers to the number of input of the Add\&Norm layer, Eq.~(\ref{eq.addnorm}). 

Whereas the BOP, the additions are taken into account as follows:
\begin{equation}\label{bop.addnorm}
   \mathrm{BOP}_{\text{Add\&Norm}} = n_xb_sb_x+(b_x+b_s+2),
\end{equation}
where $b_s$ is the bitwidth used to represent the scalar, $b_x$ is the bitwidth of the input and 2 results from two explicit additions of $x+\text{Sublayer}(x)$ and the addition of the bias.

The NABS of the Add\&Norm layer is calculated as:
\begin{equation}\label{nabs.addnorm}
   \mathrm{NABS}_{\text{Add\&Norm}} = (n_xX+1)+(b_x+b_s+2).
\end{equation}

\begin{table*}[htbp]
\caption{Summary of the three computational complexity metrics per layer (the number of real multiplications, the number of bit operations, the number of additions and bit shifts) for a zoo of neural network layers as a function of their designing hyper-parameters; the number of neurons ($n_n$),  the number of features in the input vector ($n_i$), the number of filters ($n_f$), the kernel size ($n_k$), the input time sequence size ($n_s$), the number of hidden units ($n_h$), the number of internal hidden neuron units of the reservoir ($N_r$), sparsity parameter ($s_p$), the number of output neurons ($n_o$), 
weight bitwidth ($b_w$), input bitwidth ($b_i$), activation bitwidth ($b_a$) and the number of adders required at most to represent the multiplication ($X_w$)} \label{tab:formulas}
\resizebox{\textwidth}{!}{
\begin{tabular}{|c|c|c|c|}
\hline
Network type & Real multiplications (RM)                                                                                                                           & Number of  bit-operations (BOP)                                                                                                                                                                                                                                      & Number of additions and bit shifts(NABS)                                                                                                                                                                                                                                                                       \\ \hline\hline
MLP          & $n_{n}n_{i}$                                                                                                                                        & $n_{n}n_{i}\big[b_{w} b_{i} + \text{Acc}(n_{i}, b_{w}, b_{i})\big]$                                                                                                                                                                                                  & $n_{n}n_{i}(X_{w} +1)\text{Acc}(n_{i}, b_{w}, b_{i})$                                                                                                                                                                                                                                                          \\ \hline
1D-CNN       & $n_{f} n_{i} n_{k} \cdot Output Size$                                                                                                               & \begin{tabular}[c]{@{}c@{}}$OutputSize \cdot n_{f}\text{Mult}(n_{i}n_{k}, b_{w}, b_{i})$\\ $+n_{f}\text{Acc}(n_{i}n_{k}, b_{w}, b_{i})$\end{tabular}                                                                                                                 & \begin{tabular}[c]{@{}c@{}}$OutputSize \cdot n_{f}\big[n_{i}n_{k}(X_{w} + 1 )-1\big]$ \\ $\cdot\text{Acc}(n_{i}n_{k}, b_{w}, b_{i})$\\ $+n_{f}\text{Acc}(n_{i}n_{k}, b_{w}, b_{i})$\end{tabular}                                                                                                               \\ \hline
Vanilla RNN  & $n_{s}n_{h}(n_{i} + n_{h})$                                                                                                                         & \begin{tabular}[c]{@{}c@{}}$n_{s}n_{h}\text{Mult}(n_{i}, b_{w}, b_{i})$\\ $+ n_{s}n_{h}\text{Mult}(n_{h}, b_{w}, b_{a})$\\ $+2n_{s}n_{h}\text{Acc}(n_{h}, b_{w}, b_{a})$\end{tabular}                                                                                & \begin{tabular}[c]{@{}c@{}}$n_{s}n_{h}\big[ n_{i}(X_{w} + 1) -1\big]\text{Acc}(n_{i}, b_{w}, b_{i})$\\ $+ n_{s}n_{h}\big[ n_{h}(X_{w} + 1) +1\big]\text{Acc}(n_{h}, b_{w}, b_{a})$\end{tabular}                                                                                                                \\ \hline
LSTM         & $n_{s}n_{h}(4n_{i} + 4n_{h} + 3)$                                                                                                                   & \begin{tabular}[c]{@{}c@{}}$4n_{s}n_{h}\text{Mult}(n_{i}, b_{w}, b_{i})$\\ $+ 4n_{s}n_{h}\text{Mult}(n_{h}, b_{w}, b_{a})$\\ $+ 3n_{s}n_{h}b_{a}^2$\\ $+ 9n_{s}n_{h}\text{Acc}(n_{h}, b_{w}, b_{a})$\end{tabular}                                                    & \begin{tabular}[c]{@{}c@{}}$4n_{s}n_{h}\big[n_{i}(X_{w} +1)-1\big]\text{Acc}(n_{i}, b_{w}, b_{i})$\\ $+ 4n_{s}n_{h}\big[n_{h}(X_{w} +1) +1\big]\text{Acc}(n_{h}, b_{w}, b_{a})$\\ $+ 6n_{s}n_{h}b_{a}$\end{tabular}                                                                                            \\ \hline
GRU          & $n_{s}n_{h}(3n_{i} + 3n_{h} + 3)$                                                                                                                   & \begin{tabular}[c]{@{}c@{}}$3n_{s}n_{h}\text{Mult}(n_{i}, b_{w}, b_{i})$\\ $+ 3n_{s}n_{h}\text{Mult}(n_{h}, b_{w}, b_{a})$\\ $+3n_{s}n_{h}b_{a}^2$\\ $+ 8n_{s}n_{h}\text{Acc}(n_{h}, b_{w}, b_{a})$\end{tabular}                                                     & \begin{tabular}[c]{@{}c@{}}$3n_{s}n_{h}\big[n_{i}(X_{w} + 1)-1\big]\text{Acc}(n_{i}, b_{w}, b_{i})$\\ $+ n_{s}n_{h}\big[3n_{h}(X_{w} + 1)+5\big]\text{Acc}(n_{h}, b_{w}, b_{a})$\\ $+ 6n_{s}n_{h}b_{a}$\end{tabular}                                                                                           \\ \hline
ESN          & $n_sN_{r}(n_{i} + N_{r}s_{p}+2 + n_{o})$                                                                                                            & \begin{tabular}[c]{@{}c@{}}$n_{s}N_{r}\text{Mult}(n_{i}, b_{w}, b_{i})$\\ $+ n_{s}N_{r}s_{p}\text{Mult}(N_{r}, b_{w}, b_{a})$\\ $+ n_{s}N_{r}\text{Mult}(n_{o}, b_{w}, b_{a})$\\ $+2n_{s}N_{r}b_{a}^2$\\ $+ 4n_{s}N_{r}\text{Acc}(N_{r}, b_{w}, b_{a})$\end{tabular} & \begin{tabular}[c]{@{}c@{}}$n_{s}N_{r}\big[n_{i}(X_{w} + 1) -1\big]\text{Acc}(n_{i}, b_{w}, b_{i})$\\ $+ n_{s}N_{r}\big[s_{p}(N_{r}X_{w} + N_{r} -1\big] +4 )\text{Acc}(N_{r}, b_{w}, b_{a})$\\ $+ n_{s}N_{r}\big[n_{o}(X_{w} + 1) -1\big]\text{Acc}(n_{o}, b_{w}, b_{a})$\\ $+ 4n_{s}N_{r}b_{a}$\end{tabular} \\ \hline
ResNet       & Eq.~(\ref{c.resnet})                                                                                                                                & Eq.~(\ref{bop.resnet})                                                                                                                                                                                                                                               & Eq.~(\ref{nabs.resnet})                                                                                                                                                                                                                                                                                        \\ \hline                                                                        
Transformer  & \begin{tabular}[c]{@{}c@{}}MultiHead: Eq.~(\ref{c.attention})\\ Point-wise FFN: Eq.~(\ref{c.ffn})\\ Add\&Norm: Eq.~(\ref{c.addnorm})\end{tabular}   & \begin{tabular}[c]{@{}c@{}}MultiHead: Eq.~(\ref{bop.attention})\\ Point-wise FFN: Eq.~(\ref{bop.ffn})\\ Add\&Norm: Eq.~(\ref{bop.addnorm})\end{tabular}                                                                                                              & \begin{tabular}[c]{@{}c@{}}MultiHead: Eq.~(\ref{nabs.attention})\\ Point-wise FFN: Eq.~(\ref{nabs.ffn})\\ Add\&Norm: Eq.~(\ref{nabs.addnorm})\end{tabular}                                                                                                                                    		   \\ \hline
\end{tabular}}
\label{Table:Main}
\end{table*}

\subsection{CDC block - Frequency domain equalizer}
 This FDE includes the chromatic
dispersion compensation and the recovery of the signal dispersion broadening \cite{9324921}. The FDE compensates for the dispersion by multiplying the signal by the opposite of the transfer function for dispersion. The FDE adapts its parameters on the fly according to the estimation of the built-up dispersion after the transmission. 
To have a benchmark, the computational complexity of the CDC using FDE is provided as follows \cite{9324921, spinnler2010equalizer}:
\begin{equation}\label{c.cdc}
   C_\text{{CDC}} = 4 \cdot \left(\frac{N(\log_2 N + 1)q}{N - N_{D} + 1} \right)
\end{equation}
where $N$ corresponds to the FFT size, q is the oversampling ratio and $N_{D} = q\tau_D/T$ where $\tau_D/T$ is the dispersive channel impulse response and $T$ is the symbol interval. Note that this is the RM for two polarizations, which requires four N-point FFTs and 2N complex multiplications. One N-point FFT requires $(N/2)\text{log}_2N$ complex multiplications according to the Cooley–Tukey FFT algorithm \cite{cooley1965algorithm}. Factor 4 in the expression refers to the fact that one complex multiplication can be expressed by four real ones. The term $N-N_D+1$ refers to the number of useful samples due to the overlap-save algorithm for blockwise FD filtering per polarization\cite{spinnler2010equalizer}. When considering two polarizations, the useful samples for two blocks reduced from $2N$ to $2(N-N_D+1)$. In general, the FFT size should be optimized to minimize the complexity. 

The BOP of the CDC involves complex number multiplications, therefore, first, let's define the BOP of one complex multiplication. One complex multiplication includes 4 real multiplications and 2 real additions, giving the BOP:
\begin{equation}
   \text{BOP}_\text{{complex mult}} = 4b_1 b_2 + 2(b_1+b_2+1).
\end{equation}

\begin{figure}
    \centering
    \includegraphics[scale=0.65]{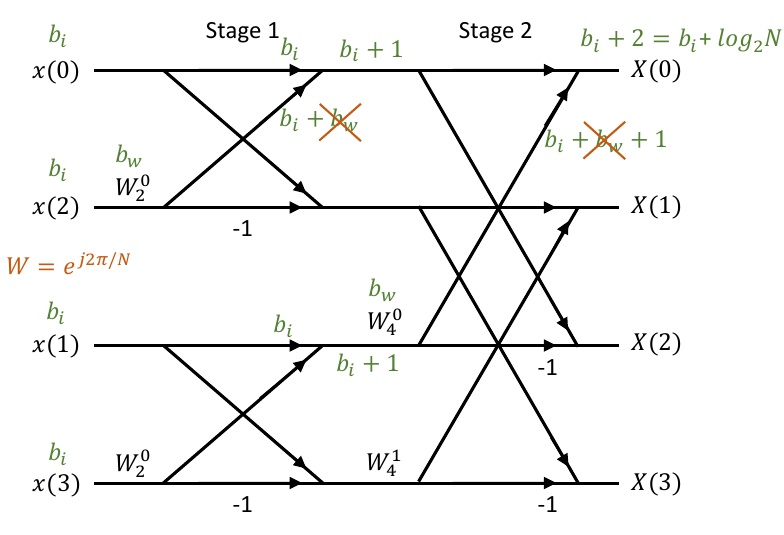}
    \caption{Butterfly computation diagram of FFT, showing the bitwidth of each step.}
    \label{fig:cooley-tukey-butterfly}
\end{figure}

\begin{figure*}[ht!]
  \centering
\begin{subfigure}{.45\textwidth}
  \centering
\includegraphics[width=0.75\textwidth]{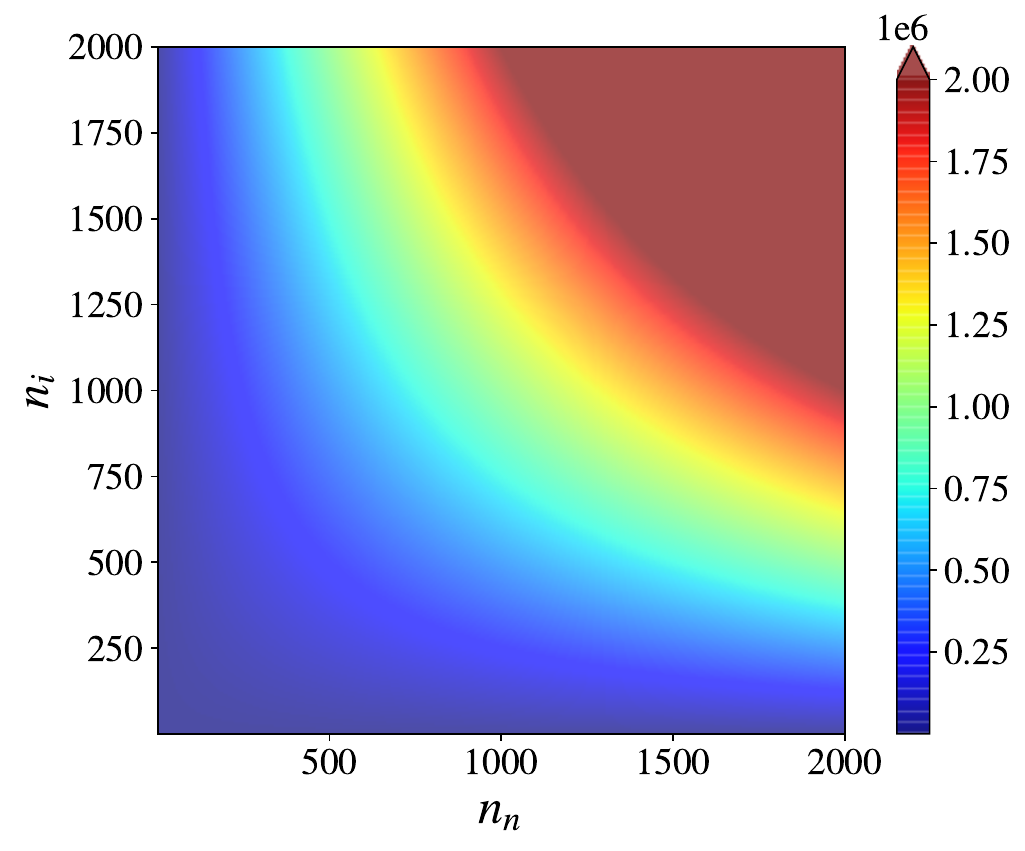}
\caption{Dense layer with parameters; number of features in the input vector $n_i$ and number of neurons in the layer $n_n$.}  \label{fig:c_feedforward_dense}
\end{subfigure}
\hfill
\begin{subfigure}{.45\textwidth}
    \centering
\includegraphics[width=0.75\textwidth]{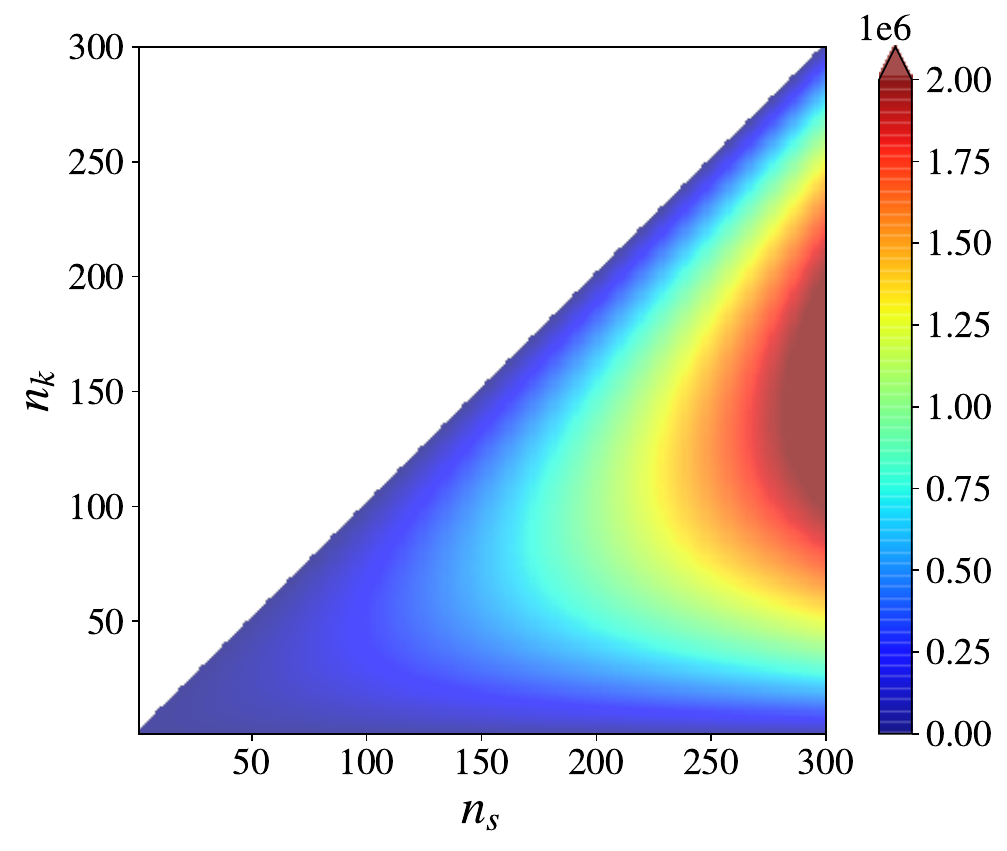}
\caption{CNN with parameters; kernel size $n_k$ and number of time steps $n_s$.} \label{fig:c_feedforward_cnn}
\end{subfigure}
\caption{Number of real multiplications (RM) of feed-forward layers.}
\label{fig:c_feedforward}
\end{figure*}

Secondly, with Cooley–Tukey FFT algorithm, one N-point FFT involves  $(N/2)\text{log}_2N$ complex multiplications and  $N\text{log}_2N$ complex additions. When doing the multiplications between the input sample and the $W$ in the FFT, we assume that because $W = e^{j2\pi/N}$, representing the unit circle, it only changes the phase of the complex values. Therefore, the multiplication between the complex number and $W$ does not change the resulting bitwidth, together with the fact that in real-world systems, there is the quantization of the result back to the desired bitwidth. The resulting bitwidth of each stage of Cooley–Tukey FFT can be illustrated in the butterfly diagram in Fig~\ref{fig:cooley-tukey-butterfly}. The BOP of one FFT can be formulated as follows:
\begin{equation}
\begin{split}
   \text{BOP}_\text{{FFT}} ={}&\frac{N}{2}\mathrm{log_2}N[4b_i b_w + 2(b_i+1)]\\
   +{}&2N\mathrm{log_2}N[b_i + \mathrm{log_2}N],  
\end{split}
\end{equation}
where $b_i$ is the bitwidth of the input sample and $b_w$ is the bitwidth of $W$ in the FFT.

Next, after the FFT, the resulting signal is multiplied by the transfer function of the filter, which contains $N$ complex multiplications per polarization of the signals. The BOP of this operation is:
\begin{equation}
\begin{split}
    \text{BOP}_\text{{TF}} ={}&N\cdot \mathrm{BOP}_{\text{complex mult}}\\
    ={}&N(4b_\text{FFT}b_\text{TF}+2(b_\text{FFT}+b_\text{TF}+1)),
\end{split}
\end{equation}
where $b_\text{TF}$ is the bitwidth of the transfer function, and $b_\text{FFT}$ is the bitwidth after the FFT. In conclusion, the BOP of the CDC contains 4 times of the BOP of the FFT and 2 times the BOP of transfer function multiplication:
The NABS of the CDC can be formulated below:
\begin{equation}
\begin{split}
    \text{BOP}_\text{{CDC}} ={}&\frac{(4\cdot \mathrm{BOP}_{\text{FFT}}+2\cdot\mathrm{BOP}_{\text{TF}})q}{2(N-N_D+1)}\\
    ={}&\frac{2Nq}{(N-N_D+1)}\Big[\text{log}_2N\Big(2b_ib_w+(b_i+1)\\
    +{}&2(b_i +\text{log}_2N)\Big)\\
    +{}&2b_\text{FFT}b_\text{TF}+(b_\text{FFT}+b_\text{TF}+1)\Big],
\end{split}
\end{equation}

The NABS of the CDC can be formulated below:
\begin{equation}
\begin{split}
    \text{NABS}_\text{{CDC}} 
    ={}&\frac{2Nq}{(N-N_D+1)}\Big[\text{log}_2N\Big((2X_w+1)(b_i+1)\\
    +{}&2(b_i +\text{log}_2N)\Big)\\
    +{}&(2X_\text{TF}+1)(b_\text{FFT}+b_\text{TF}+1)\Big],
\end{split}
\end{equation}
where $X_\text{TF}$ represents the number of adders we need at most to perform the multiplication of the transfer function, see more explanation in Section~\ref{sec.dense}.

\begin{figure*}[ht!]
  \centering
\begin{subfigure}{.45\textwidth}
  \centering
\includegraphics[width=0.85\textwidth]{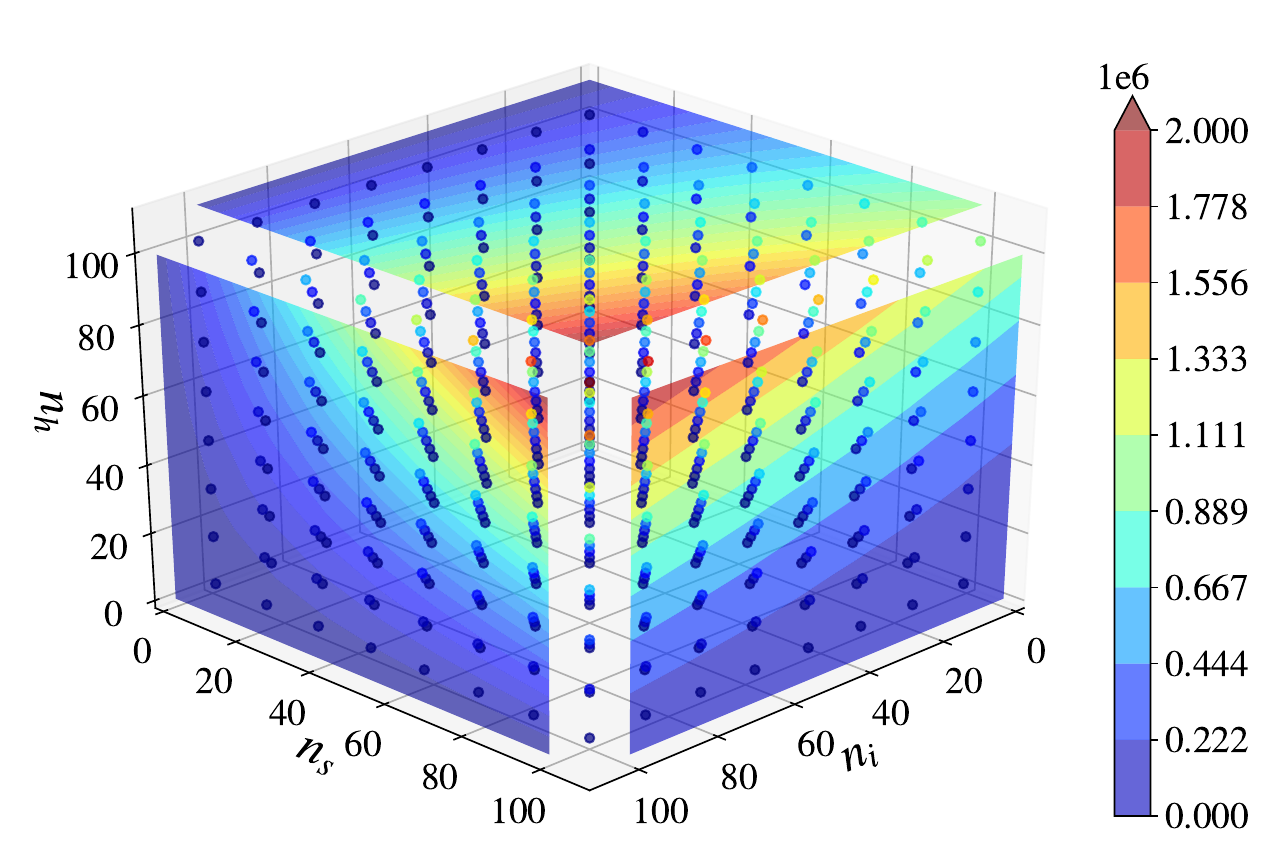}
\caption{Vanilla RNN.} \label{fig:c_rnn_cube} 
\end{subfigure}
\hfill
\begin{subfigure}{.45\textwidth}
    \centering
\includegraphics[width=0.85\textwidth]{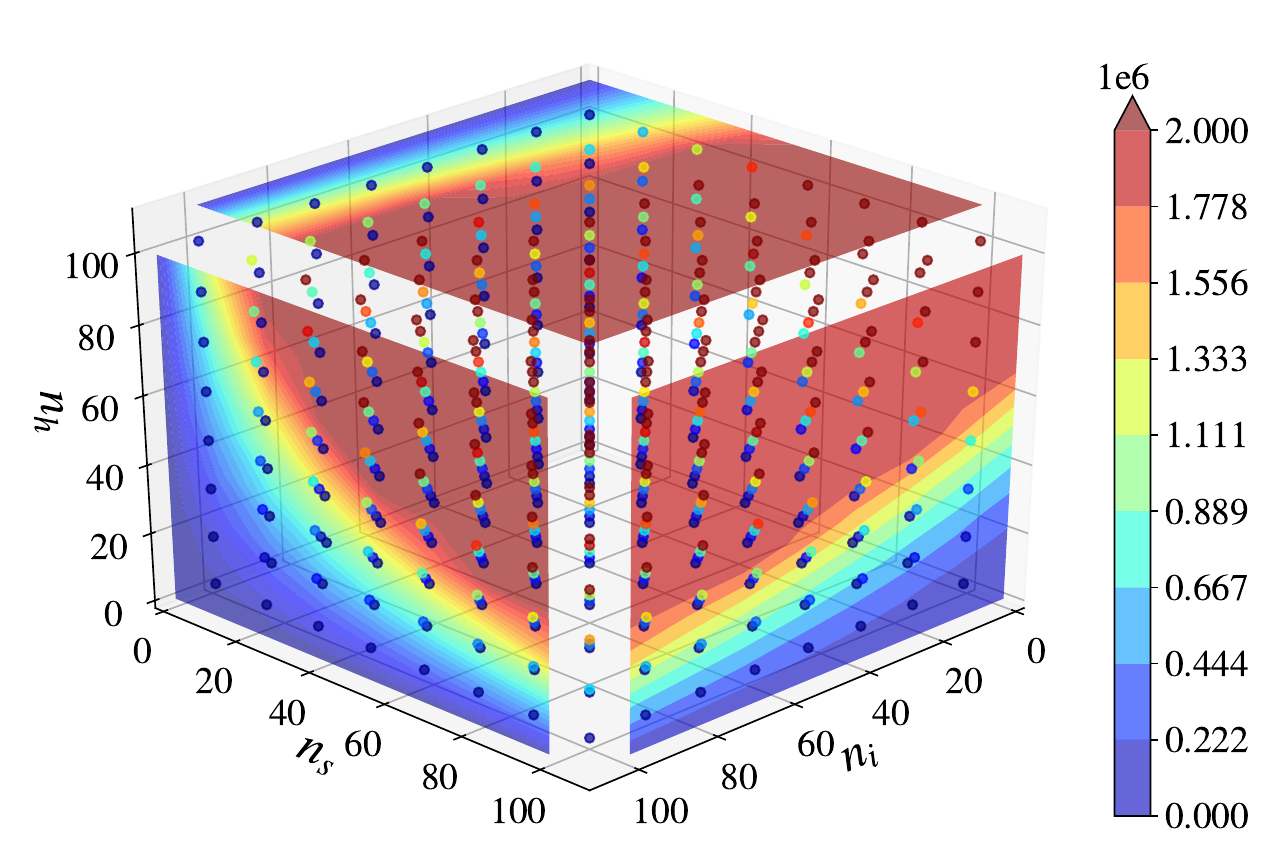}
\caption{LSTM.}\label{fig:c_lstm_cube}
\end{subfigure}
\medskip
\begin{subfigure}{.45\textwidth}
    \centering
\includegraphics[width=0.85\textwidth]{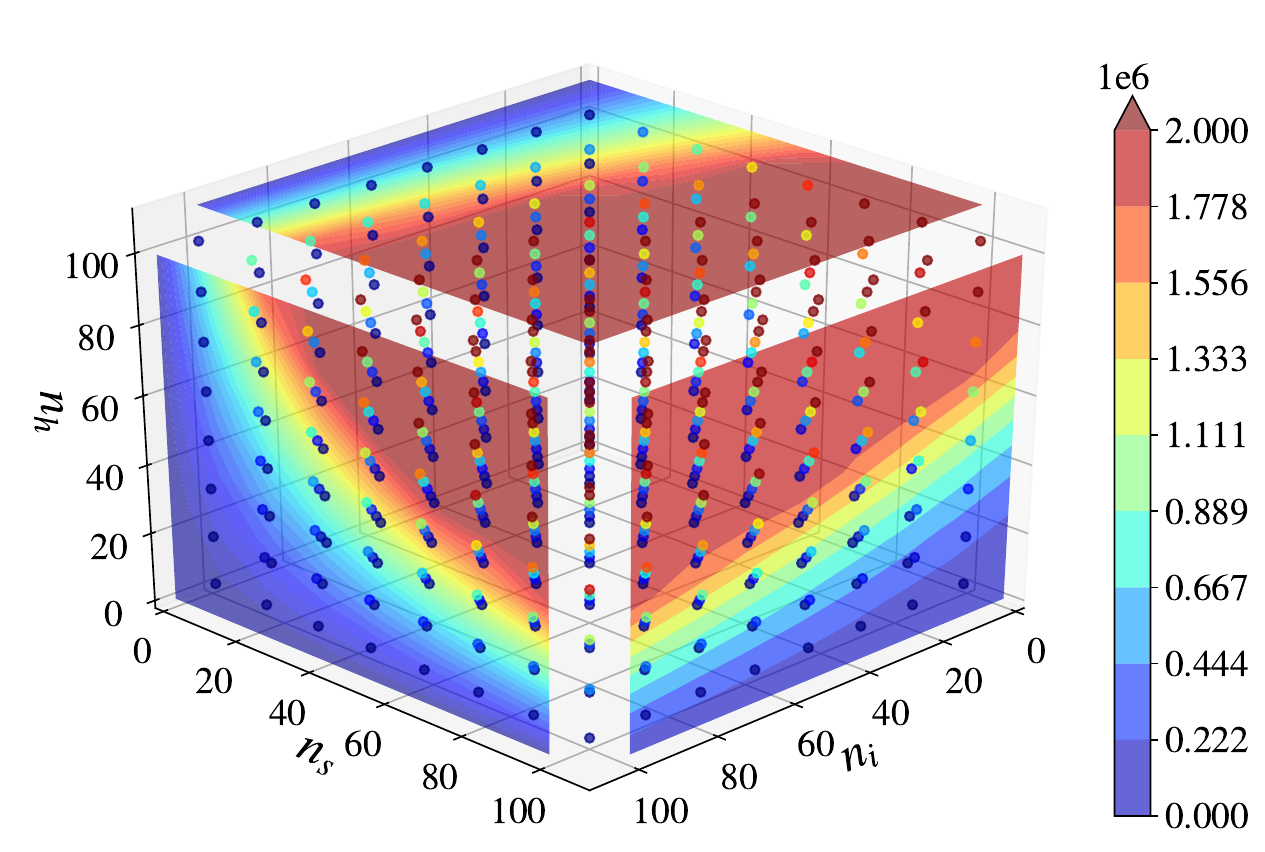}
\caption{GRU.}\label{fig:c_gru_cube}
\end{subfigure}
\hfill
\begin{subfigure}{.45\textwidth}
  \centering
\includegraphics[width=0.85\textwidth]{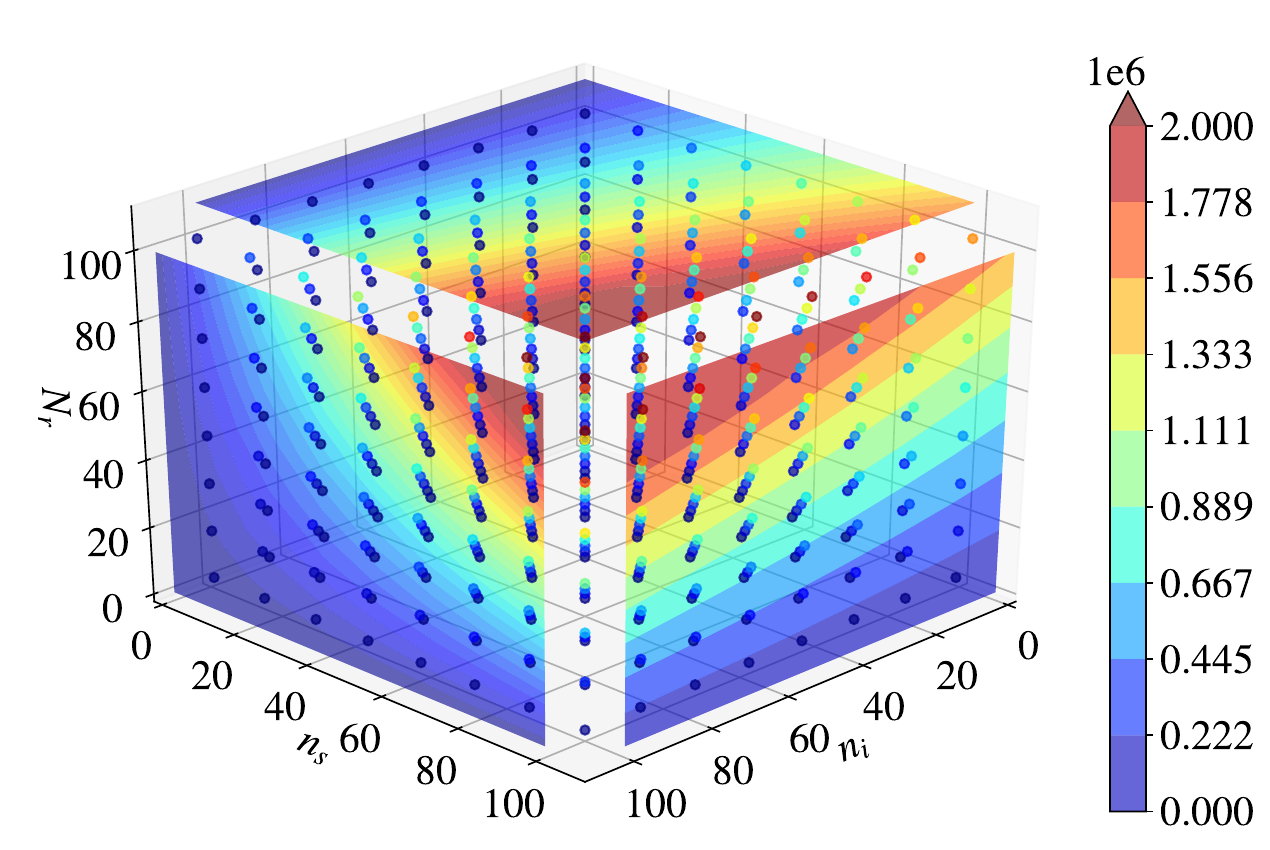}
\caption{ESN.}  \label{fig:c_esn_cube}
\end{subfigure}
\caption{Number of real multiplications (RM) of recurrent layers with respect to different values of the number of features in the input vector $n_i$, the number of time steps $n_s$ and the number of hidden units $n_h$ or $N_r$ in ESN.}
\label{fig:c_recurrent_cube}
\end{figure*}

\subsection{Digital back propagation (DBP)}
This DBP technique utilizes the symmetric split-step Fourier approach for its implementation. Here, the parameters of the linear filter and the nonlinear operators are fine-tuned to minimize the equalized BER. It was assumed that the nonlinear step would remain entirely static. When considering a single channel, the computational demand of the DBP method, measured by the necessary RM per transmitted symbol, can be approximated as \cite{9324921,napoli2014reduced}:

\begin{equation}\label{c.dbp}
   C_\text{{DBP-1CH}} = 4qN_{\text{Sp}}N_{\text{StSP}}\left(\frac{N(\log_2 N + 1)}{N - N_{Dq} + 1} + 1\right)
\end{equation}
where $N_{Sp}$ refers to the total number of spans, $N_{STpS}$ is the number of propagation steps per span, and q is the oversampling factor. The RM of DBP of each step consists of the linear part which is the RM of the CDC and the nonlinear part which is the addition of one RM refers to the multiplication of a nonlinear term. 

The BOP of the DBP can be calculated, similarly to the RM of the CDC, as follows:
\begin{equation} \label{BOP.dbp}
\begin{split}
\mathrm{BOP}_{\text{DBP}}
={}&qN_{\text{Sp}}N_{\text{StSP}}
\biggl[
\frac{2N}{(N-N_{D}+1)} \biggl(2b_\text{FFT}b_\text{TF} \\
+{}&(b_\text{FFT}b_\text{TF} + 1)\\
+{}&\log_{2}N(2b_{i}b_{w} + (b_{i}+1) + 2(b_i+\log_{2}N)\biggr)\\
+{}&4b_\text{CDC}b_\text{NL} + 2(b_\text{CDC}+1)
\biggr],\\
\end{split}
\end{equation}
where $b_{i}$ corresponds to the bitwidth of the input symbol, $b_{w}$ is the bitwidth of the $W$ value as described in the CDC. Here, we assume that multiplying the input with $W$ does not change the resulting bitwidth as the $W$ refers to the phase change as described in the CDC. $b_\text{CDC}$ refers to the resulting bitwidth after the linear step (CDC), which usually is quantized to a certain number of bits. Lastly, $b_text{NL}$ is the bitwidth of the nonlinear term that applies a nonlinear phase shift to the signal.

The NABS of DBP can be found as:
\begin{equation} \label{nabs.dbp}
\begin{split}
\mathrm{NABS}_{\text{DBP}}
={}&qN_{\text{Sp}}N_{\text{StSP}}
\biggl[
\frac{2N}{N-N_{D}+1} \biggl((2X+1)(b_\text{FFT}b_\text{TF} + 1)\\
+{}&\log_{2}N((2X+1)(b_{i}+1) + 2(b_i + \log_{2}N)\biggr)\\
+{}&(4X+2)(b_\text{CDC}+1)
\biggr].\\
\end{split}
\end{equation}
The multiplication is changed to the number of additions for the NABS.

\section{Comparative analysis of the complexities for each NN structure}\label{Sec:results}
The comparison of the complexity in terms of RM is illustrated in Fig.~\ref{fig:c_feedforward} for feed-forward NNs and in Fig.~\ref{fig:c_recurrent_cube} for recurrent networks. In feed-forward networks, we first address the computational complexity of a dense layer. To reach over $2\times10^6$ real multiplications, which we use as a threshold (highlighted by a maroon color), we can have up to around 1500 input features ($n_i$) and 1500 neurons ($n_n$) which is a high value for a single dense layer. The $n_n n_i$ term in Eq.~(\ref{c.dense}) forms a hyperbolic curve, as can be seen in Fig. \ref{fig:c_feedforward_dense}. For the 1D-convolutional layer, as predicted by Eq.~\ref{c.cnn}, we reach a high complexity (maroon) region using fewer input features than the dense layer case because now the complexity growth depends on more than 2 variables (e.g. to reach the complexity threshold, the number of time steps $n_s$ can be set to 275 with the kernel size $n_k$ equal to 150, and we fixed $n_i = 100$, $n_o = 1$, $padding = 0$, $dilation = 1$, and $stride = 1$). According to the exemplary chosen parameters above, the $n_k \le n_s$ condition derived from Eq.~(\ref{outputwidth.cnn}) must be satisfied to obtain at least the output size equal to 1; therefore, in Fig.~\ref{fig:c_feedforward_cnn}, the white region corresponds to unavailable output and the heat-map has a hyperbolic form because only $n_s$ and $n_k$ parameters vary and the other parameters are kept constant.

\begin{figure*}[!ht]
\begin{subfigure}[c]{0.32\linewidth}
\centering
\includegraphics[width=\textwidth]{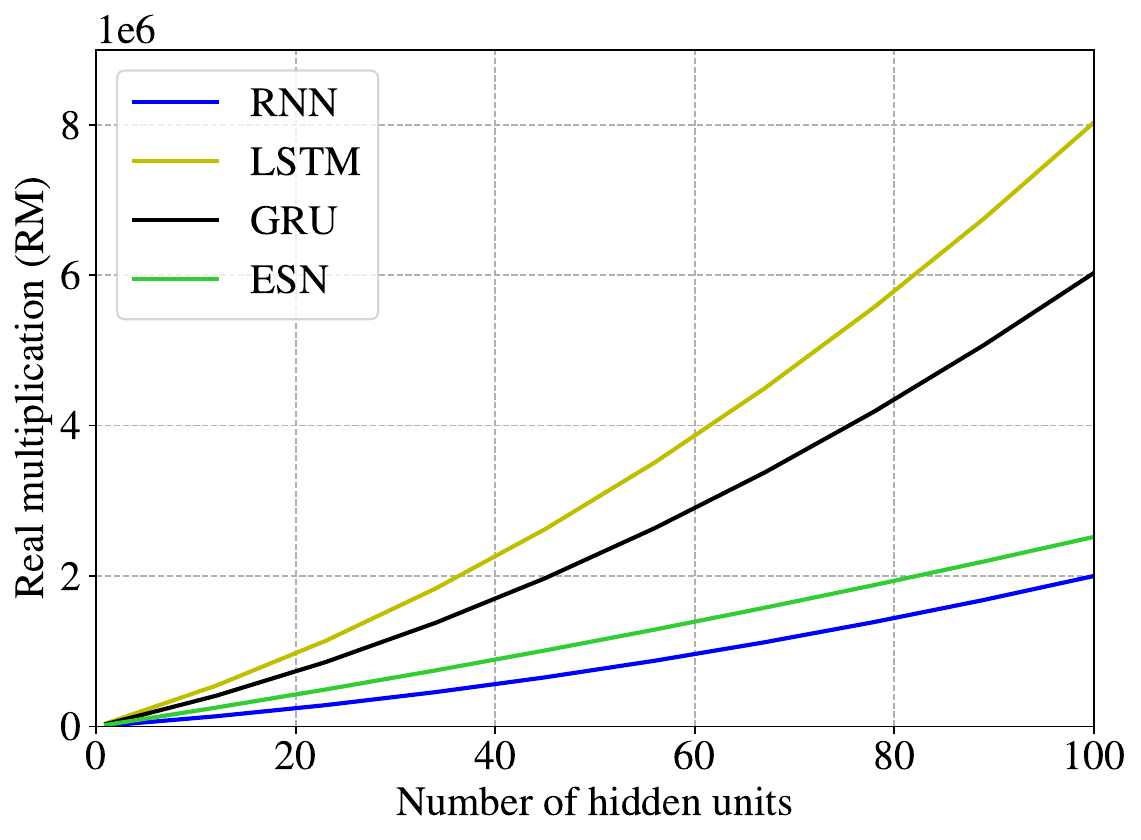}
\caption{}
    \label{fig:complxity.compare}
\end{subfigure}\hfill
\begin{subfigure}[c]{0.32\linewidth}
\centering
\includegraphics[width=\textwidth]{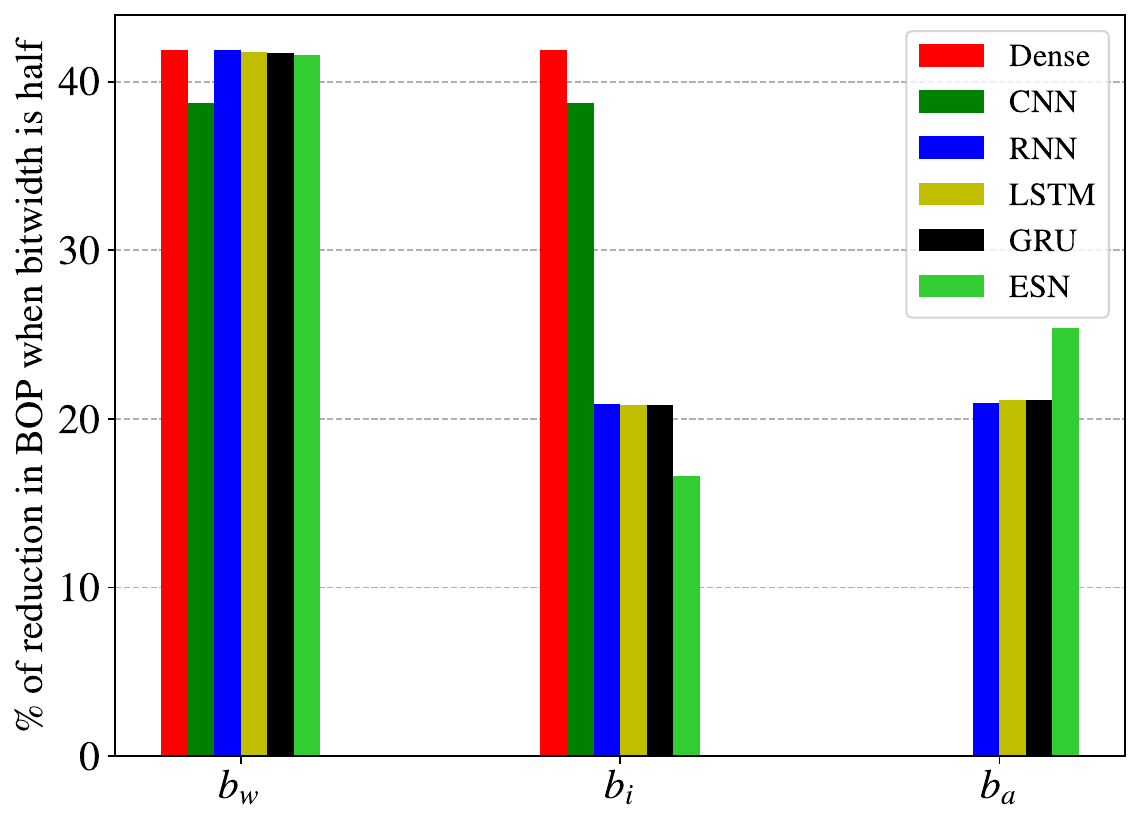}
\caption{}
\label{fig:BOPs_reduction}
\end{subfigure}\hfill
\begin{subfigure}[c]{0.32\linewidth}
\centering
\includegraphics[width=\textwidth]{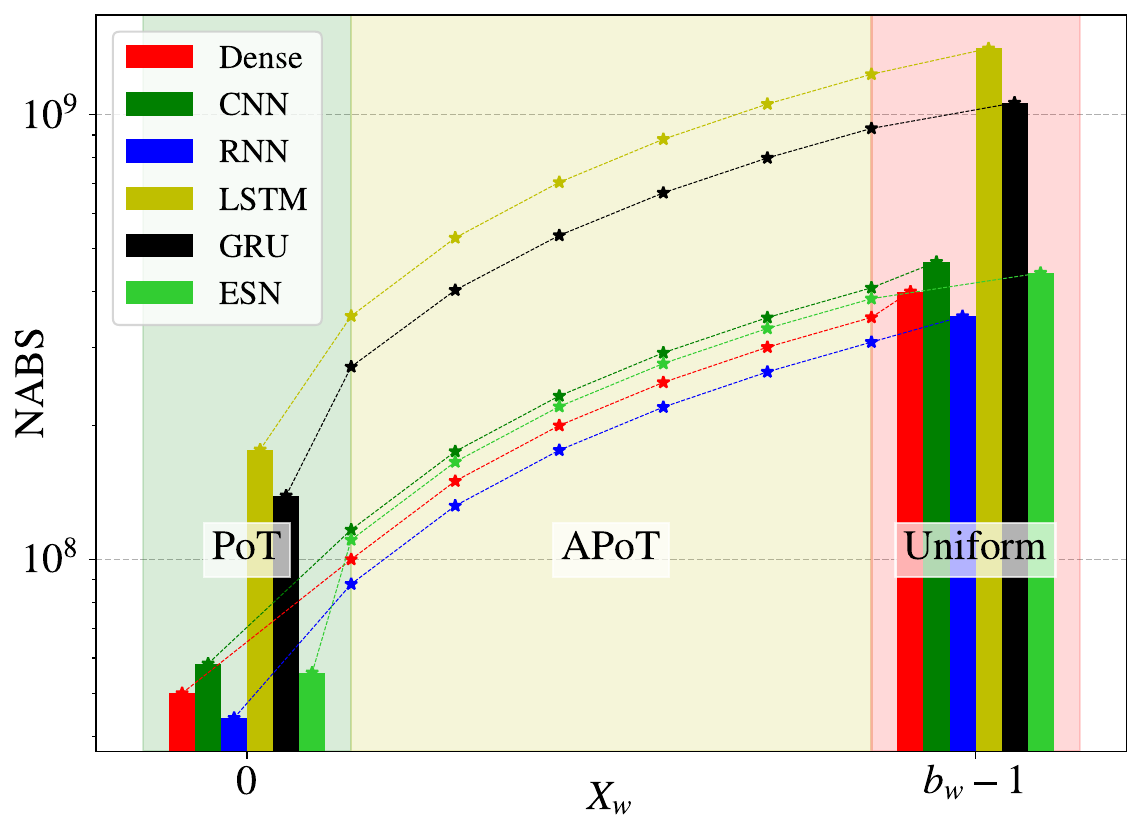}
\caption{}
\label{fig:nabs}
\end{subfigure}
\caption{(a) Complexity comparison between recurrent-based networks when $n_s$ = 100, $n_i$ = 100 for all networks and $n_o$ = 100, $s_p$ = 0.5 for ESN; (b) Reduction of BOP in percentage when reducing bitwidth of each parameter; weight bitwidth $b_w$, input bitwidth $b_i$, and activation bitwidth $b_a$ by half or from 8 bits to 4 bits; (c) Comparison of the number of additions and bit shifts (NABS) with different quantization techniques and different network types, assume $b_w = 8$. Note that $X_w$ is the number of adders required to represent a multiplier. }
\label{fig:MainResults}
\end{figure*}

The RNN-based networks apparently have higher complexity than the feed-forward NNs. The vanilla RNN in Fig. \ref{fig:c_rnn_cube} shows the least complexity among the RNN-based networks studied, while the LSTM's complexity growth is the fastest, which can be seen from the size of the maroon area in Fig. \ref{fig:c_lstm_cube}. The GRU in Fig. \ref{fig:c_gru_cube} shows slightly lower complexity than the LSTM because it has a lower number of gates in its architecture. If we look at the equations of GRU and LSTM in Table \ref{tab:formulas}, the LSTM has a multiplier of 4 for the $n_i$ and $n_h$, whereas for the GRU the multiplier is 3. The ESN with fixed $n_o = 100$ and $s_p = 0.5$ in Fig. \ref{fig:c_esn_cube} has higher complexity than the vanilla RNN, but less complexity than the GRU, because the ESN by design has a less complex architecture due to the use of the reservoir \cite{lopez2020comparison}. For all RNN-based networks, we readily infer that the number of hidden units $n_h$, or $N_r$ for the ESN, plays the most crucial role in defining the layer's computational complexity in terms of RM metric. In Fig.\ref{fig:c_recurrent_cube}, we observe that the top face of all cubes which corresponds to the highest number of $n_h$ ($n_h = 100$) has the largest maroon areas. In terms of the effect on the complexity behavior, the second most important quantity is the number of time steps $n_s$; we can see in Fig.~\ref{fig:c_recurrent_cube} that the right face of the cubes, referring to the highest number of $n_s$ ($n_s = 100$), has the second-largest maroon areas. Finally, the dimensions of the input vector $n_i$ have the least impact on the RM, as shown by the left face ($n_i = 100$) of all cubes in Fig.~\ref{fig:c_recurrent_cube} with the smallest maroon areas compared to other faces. See Eqs.~(\ref{c.dense}), (\ref{c.cnn}), (\ref{c.rnn}), (\ref{c.lstm}), (\ref{c.gru}), and (\ref{c.esn}) for the exact  dependencies.

Furthermore, to highlight the computational complexity trend over those different recurrent layers, we plotted the RM versus the number of hidden units ($n_h$ for vanilla RNN, LSTM, and GRU, or $N_r$ in ESN) in a scenario where all other hyperparameters are constant. Fig.~\ref{fig:complxity.compare} depicts the result of this analysis when $n_s = 100$, $n_i = 100$ for all networks, and $n_o = 100$, $s_p = 0.5$ for the ESN.  By considering those parameters as fixed, the complexity of all four recurrent layers scales quadratically with the number of hidden units ($n_h^2$), which is traditionally interpreted as having the same $O(n^2)$. However, Fig.~\ref{fig:complxity.compare} brings an important fact that there are significant differences in the computational complexity in terms of RM between all four recurrent layers. Ultimately, this means that the Big-$O$ notation is not sensitive enough to assess the complexity of the NNs in digital signal processing. We can spot that the LSTM complexity escalates the fastest followed by the GRU, ESN, and RNN, respectively. These differences result mainly from the scaling terms on the $n_h^2$ of each RM complexity expression for these layers. Note that for the ESN (Eq.~(\ref{c.esn})), the complexity increases more steadily, as far as $N_r$ is multiplied by the sparsity parameter $s_p$. Moreover, as noted in Sec.~\ref{sec.esn}, the reservoir can be implemented in the optical domain, so the complexity can be reduced further at the expense of performance trade-off.

We notice that not only the hyper-parameters like $n_h$ and $n_s$, affect the computational complexity, but also the bitwidth or precision of each parameter can impact the complexity when we quantify it in terms of the BOP. The effects produced by the bitwidth value of weight ($b_w$), input ($b_i$), and activation ($b_a$) are examined in this study. The full-precision or 32-bit precision can be considered over-redundant because 8-bit or less is often enough to provide comparable performance, as stated by many works in the field \cite{yu2019any, banner2018scalable, wang2019haq, hubara2021accurate}. Here, we did not provide the study of the BOP versus different hyper-parameters and bitwidth, because we believe that this is a straightforward analysis that approximately follows what we already studied with the RM metric. Instead, we focus on answering the following question: which variable bitwidth ($b_w$, $b_i$, or $b_a$) produces the highest saving in the BOP complexity when its precision is reduced? This question can guide the design of a low-complexity NN structure by identifying which parameter of the NN is the key to producing the higher saving in complexity. To address this question, we compare the reduction of the BOP when using 8-bit precision versus the 4-bit precision for each parameter ($b_w$, $b_i$, and $b_a$) in different network types; the results of the comparison are shown in Fig.~\ref{fig:BOPs_reduction}. The bitwidth of the weight matrix $b_w$ is the most significant parameter to consider when trying to reduce the layer's complexity, as the BOP is decreased by around 40\% for all network types when we halved the precision of $b_w$. For the dense and 1D-convolutional layers, the precision for input $b_i$ is as important as the $b_w$, while reducing the bitwidth of the bias vector $b_a$ does not have a noticeable impact on the BOP. In the RNN-based networks, converting from 8-bit to 4-bit precision for $b_i$ and $b_a$ shows a nearly equivalent reduction in the BOP, except for the ESN case, where decreasing the $b_a$ precision results in more reduction in the BOP than when we reduce the $b_i$ precision.  

Lastly, we analyze the NABS metric considering various quantization techniques: uniform, PoT, and APoT quantization, as described in Sec.~\ref{sec.dense}. Note that each technique needs a different number of shifts and adders to perform the multiplication. As mentioned before, the shifts incur no extra cost in hardware implementation; therefore, we focus on evaluating the NABS metric of each layer for certain quantization techniques versus the number of adders required at most to perform the multiplication, denoting it as $X$. More specifically, if the weight matrix has $b_w$ as its bitwidth and the uniform quantization is utilized, the number of adders required at most ($X_w$) is equal $b_w-1$. In the case of PoT, $X_w = 0$, and for APoT, $X_w$ varies between 1 and $b_w-2$, in this case. Fig.~\ref{fig:nabs} shows the NABS versus $X_w$ analysis when considering that: $b_w,b_i,b_a=8$ for all networks, $n_i = 1000$ and $n_n = 2000$ for a dense layer, $n_i = 100$, $n_s = 300$, $n_o = 1$, padding $= 0$, dilation $= 1$, stride $= 1$  and $n_k = 100$ for a 1D-convolutional layer, $n_i = 100$, $n_s = 100$ and $n_h = 100$ for all RNN-based networks, and $n_o = 100$, $s_p = 0.5$ for the ESN. 

As shown in Fig.~\ref{fig:nabs}, for all types of networks, when the PoT quantization is used, the NABS can drop around 8 times lower compared to the NABS when using the uniform quantization. Since APoT is a quantization scheme represented by a sum of PoT terms, APoT provides a smooth transition between PoT and uniform quantization. In various works, PoT was claimed to have very low complexity because the multiplications are replaced by just shifts \cite{marchesi1993fast,chang2020msp,przewlocka2022power}. However, when we consider that the multiplication in the uniform quantization can be represented by shifts and adders, and we have a fair metric like NABS to compare between different quantization techniques, the NABS when applying PoT is only around an order of magnitude lower than the NABS when using the uniform quantization. To be more specific, even though PoT converts all multipliers into bit shifters, we still have a number of adders coming from the sum operations that are not related to the multipliers, but they are key for the operational structure of the NN layers. Therefore, the NABS metric can provide a reliable assessment of the computational complexity of NNs before their implementation in hardware, where the NLG will be the ultimate metric. Note that we intentionally increased the values of the hyper-parameters for the feed-forward NNs in order to compare them in the same graph as the RNN-based networks. For the complexity with regard to NABS, the LSTM apparently needs the highest number of shifts and adders. In conclusion, the three matrices of complexity: the RM, the BOP, and the NABS, have the same trend, meaning that the LSTM requires the most computational resources, followed by the GRU. However, the complexity depends on the particular NN design and can be reduced if we can tolerate a more accurate trade-off: varying the values of hyper-parameters can affect the accuracy, but, simultaneously, work in favor of reducing the computational complexity.

\begin{table*}[ht!]
\centering
\begin{tabular}{c|c|c|c|c|c|c|c}
Model Type & Q-factor       & \begin{tabular}[c]{@{}c@{}}No. of Trainable \\ Parameters\end{tabular} & RMpS                         & BOPpS                         & NABSpS                         & \begin{tabular}[c]{@{}c@{}}CPU Inference Time\\ per Window\end{tabular} & \begin{tabular}[c]{@{}c@{}}GPU Inference Time\\ per Window\end{tabular} \\ \hline \hline
biLSTM+CNN (Teacher) & \textbf{10.66} & \textbf{104,407}                                                       & \textbf{1.28$\times10^{5}$} & \textbf{1.66$\times10^{8}$} & \textbf{3.19$\times10^{8}$} & 5.78$\times10^{-3}$                                                     & 7.65$\times10^{-3}$                                                     \\
1D-CNN (Student)    & 10.19          & 293,390                                                                & 3.72$\times10^{5}$         & 8$\times10^{8}$          & 1.27$\times10^{9}$          & \textbf{5.12$\times10^{-3}$}                                            & \textbf{3.87$\times10^{-4}$}                                           
\end{tabular}
\caption{Summary of the performance versus complexity of two NN-based equalizers after applying KD  (biLSTM+CNN as a Teacher model and 1D-CNN as a Student model), where the bitwidth b$_i$ = 64, b$_w$ = 32, and b$_a$ = 32. The RM, BOP and NABS are reported \"per equalized symbol\".}
\label{tab:KD_models}
\end{table*}

\section{A Practical Study on the Benefits of Complexity Reduction in NN-based Equalizer}\label{Sec:case-study}
\begin{figure}[ht!]
    \centering
    \includegraphics[scale=0.4]{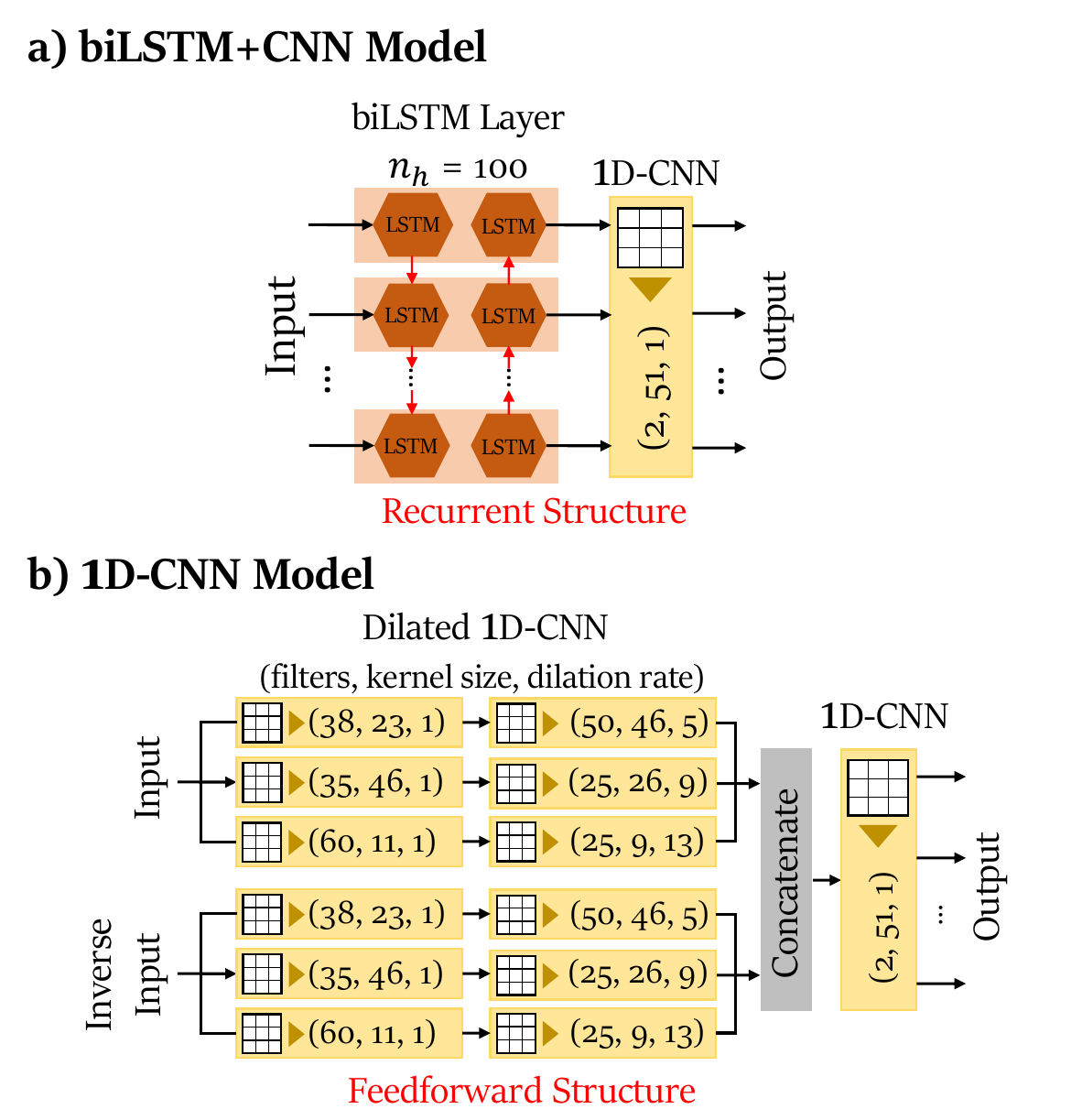}
    \caption{NN-based equalizer structure considered: a) biLSTM+CNN and b) 1D-CNN models.}
    \label{fig:nn_structures}
\end{figure}

This section presents a practical study of the impact of the complexity reduction techniques on the optical performance of a numerically simulated single 64~QAM 30~GBd dual-polarization channel transmitted along 20$\times$50km of SSMF. The NN structures considered in this paper are a biLSTM+CNN model and a 1D-CNN model. Both models take an input window of 221 symbols to recover 171 output symbols. The biLSTM model, shown in Fig.~\ref{fig:nn_structures}, contains 100 hidden units, as explained in~\cite{freire2023reducing}. For the 1D-CNN model,  the dilated CNN was applied, as explained in detail in Ref.~\cite{srivallapanondh2023parallelization}, and the NN parameters are presented in Fig.~\ref{fig:nn_structures}b.

\begin{figure}[hb!]
    \centering
\begin{tikzpicture}[scale=0.97]
  \begin{axis}[
    ybar=5*\pgflinewidth,
    symbolic x coords={2, 3, 4, 5, 6},
    xlabel = {Bitwidth of Weights [bits]},
    bar width=0.15cm,
    enlarge x limits=0.13,
        ymin=0,ymax=12,
    xtick=data,
    ymajorgrids = true,
    ylabel = {\textcolor{black}{Q-factor [dB]}} ,
        legend style={at={(0.5,1.115)},
anchor=north,legend columns=-1},
    extra y ticks = 10.7, 
        extra y tick labels={},
        extra y tick style={grid=major,major grid style={thick, dashed,draw=red}}
    ]

    \addplot[style={green!80,fill=green!80}]
          coordinates {(2,5.2)(3,6.39)(4,7.9)(5,9.46) (6,10.1)    };
    \addplot[style={orange!80,fill=orange!80,mark=none}]
          coordinates {(2,5.2)(3,6.39)(4,7.6)(5,7.65) (6,7.79) };
    \addplot[style={blue!80,fill=blue!80,mark=none}]
          coordinates {(2,5.2)(3,6.39)(4,8.33)(5,9.24) (6,9.24)};
    \addplot[style={red!70,fill=red!70,mark=none}]
          coordinates {(2, 8.57) (3, 9.37) (4, 9.93) (5, 10.347)  (6, 10.53)};
\legend{Uniform, PoT, APoT-2 terms, W.C.}
  \end{axis}

      \draw[dashed, draw=black] (0,4.1) -- (6.88,4.1);

      \draw[dashed, draw=black] (0,2.55) -- (6.88,2.55);
      
      \node[text width=4.5cm, red] at (2.37,5.29) 
    {Ref-Original (NN)};

          \node[text width=4.5cm, black] at (2.37,4.32) 
    {DBP};

          \node[text width=4.5cm, black] at (2.37,2.8) 
    {CDC};
    
\end{tikzpicture}
    \caption{Q-factor vs. bit-width when employing  Uniform Quantization, PoT, APoT with two adders, and W.C. The model w/o quantization is shown as a benchmark together with the CDC and 1 StPS DBP, at 2 dBm launch power.}
    \label{fig:results}
\end{figure}
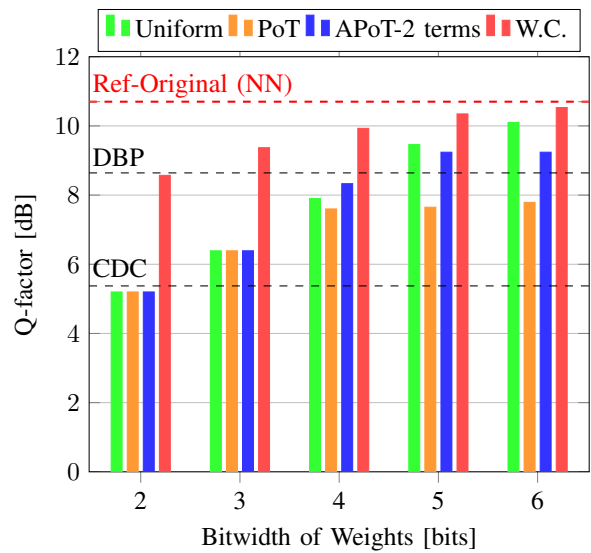

\subsection{Analysis of performance of quantization techniques}
We present an example of how such metrics can be powerful in reducing the computational complexity of NN-based equalizers when compared with traditional DSP equalizers such as the CDC block and the DBP method. In this analysis, the aforementioned biLSTM+CNN model was evaluated. The launch power was set to 2 dBm, which leads to the highest Q-factor after nonlinear equalization.


In this study, the performance of NN is evaluated when quantization-aware training (QAT) is implemented to determine whether training can mitigate the error introduced by the low bit precision of the NN weights. Fig.~\ref{fig:results} summarizes the Q-factor as a function of bitwidth, considering quantization with different schemes: Uniform\cite{goyalfixed}, Power-of-Two (PoT) \cite{zhouincremental} and Additive Power-of-Two (APoT) with 2 terms \cite{liadditive} and W.C. \cite{han2015deep}\footnote{We assume that the weights only are quantized while the inputs are still considered to be float32.}. The original model without quantization, the standard 1 Step-per-span (StPS) digital backpropagation (DBP), and Chromatic Dispersion Compensation (CDC) are used as a benchmark. As expected, as the complexity is reduced (lower precision/bit-width), the NN tends to perform worse in all the complexity reduction schemes considered.

The analysis of Fig.~\ref{fig:results} shows that the W.C. approach outperforms the other quantization techniques due to its ability to learn the optimal alphabet for each part of the NN structure instead of using a static quantized alphabet like uniform, PoT, and APoT methods. Consequently, the performance of the equalizers is less affected. Notably, with a 6-bit alphabet size of $2^6$ clusters, the W.C. technique achieves similar performance to the original model, while a 2-bit alphabet size performs similarly to the 1 StPS DBP. Regarding the trade-off between optical performance and computational complexity, the APoT approach with two terms may be a feasible option, as it only requires one adder for each multiplication, as detailed in\cite{koike2021zero}.

We note that the use of QAT can result in an unstable training process\footnote{The training phase of a quantized model may encounter obstacles associated with learning, such as the exploration versus exploitation trade-off, hyperparameter sensitivity, loss leading to NaN, or gradient problems.}, which requires continuous training monitoring.  Moreover, when considering smaller bit levels, we recommend implementing a gradual quantization approach in which the precision is gradually decreased during the training process while optimizing the learning rate and batch size.

\subsection{Complexity comparison between two NN-based equalizers}
In this section, we evaluate the performance versus computational complexity of the NN-based equalizers after applying KD framework to train the 1D-CNN model (Student) with the knowledge of the biLSTM+CNN (Teacher) as in Ref.~\cite{srivallapanondh2023parallelization}. The computational complexity in terms of the number of trainable parameters, RM, BOPs, NABS, and inference latency are compared. As mentioned earlier, KD in general context is used to reduce the computational complexity in terms of the NN parameters, however, in Ref.~\cite{srivallapanondh2023parallelization}, the KD framework is used to recast the NN structure from biLSTM-based (recurrent-based) to a feedforward-based equalizer to allow parallelization in processing. In this case, KD was applied to focus on enabling parallelization and reducing the inference latency rather than reducing the RM, BOPs, or NABS.

\begin{figure}[ht!]
    \centering
\begin{tikzpicture}[scale=0.97]
  \begin{axis}[
    xtick={2,4,6,8,10,12,14,16},
    extra x ticks={3,5,7,9,11,13,15},
    extra x tick labels={},
    xlabel = {Bitwidth of Weights [bits]},
    enlarge x limits=0,
    ymin=0,
    xtick=data,
    ymajorgrids = true,
    xmajorgrids = true,
    ylabel = {\textcolor{blue}{BOPs}, \textcolor{red}{NABS} per equalized symbol} ,
    legend style={at={(0.01,0.99)}, nodes={scale=0.87, transform shape},anchor=north west, legend cell align=left,fill=white, fill opacity=0.6, draw opacity=1,text opacity=1},
    grid style={dashed}]
    ] 
    \addplot[color=blue, mark=diamond, very thick]
          coordinates {( 2 , 4983441 ) ( 3 , 6331330 ) ( 4 , 7679220 ) ( 5 , 9027109 ) ( 6 , 10374998 ) ( 7 , 11722887 ) ( 8 , 13070777 ) ( 9 , 14418666 ) ( 10 , 15766555 ) ( 11 , 17114444 ) ( 12 , 18462334 ) ( 13 , 19810223 ) ( 14 , 21158112 ) ( 15 , 22506001 ) ( 16 , 23853890 )};
          \addlegendentry{BOPs of biLSTM+CNN};
    \addplot[color=blue, mark=square, very thick, dash pattern={on 7pt off 3pt}]
          coordinates {( 2 , 22554905 )( 3 , 28887081 )( 4 , 35219257 )( 5 , 41551433 )( 6 , 47883609 )( 7 , 54215785 )( 8 , 60547961 )( 9 , 66880137 )( 10 , 73212314 )( 11 , 79544490 )( 12 , 85876666 )( 13 , 92208842 )( 14 , 98541018 )( 15 , 104873194 )( 16 , 111205370 )};
          \addlegendentry{BOPs of 1D-CNN};
              \addplot[color=red, mark=diamond, very thick]
          coordinates {( 2 , 4995590 ) ( 3 , 7872544 ) ( 4 , 11005354 ) ( 5 , 14394019 ) ( 6 , 18038538 ) ( 7 , 21938913 ) ( 8 , 26095143 ) ( 9 , 30507227 ) ( 10 , 35175167 ) ( 11 , 40098961 ) ( 12 , 45278611 ) ( 13 , 50714115 ) ( 14 , 56405475 ) ( 15 , 62352689 ) ( 16 , 68555758 ) };
          \addlegendentry{NABS of biLSTM+CNN};
    \addplot[color=red, mark=square, very thick, dash pattern={on 7pt off 3pt}]
          coordinates {( 2 , 21284654 )( 3 , 33051873 )( 4 , 45564124 )( 5 , 58821408 )( 6 , 72823725 )( 7 , 87571075 )( 8 , 103063457 )( 9 , 119300871 )( 10 , 136283319 )( 11 , 154010799 )( 12 , 172483312 )( 13 , 191700857 )( 14 , 211663435 )( 15 , 232371046 )( 16 , 253823689 ) };
          \addlegendentry{NABS of 1D-CNN};
  \end{axis}
\end{tikzpicture}
    \caption{BOPs and NABS per equalized symbol as a function of bitwidth of weights of biLSTM+CNN and 1D-CNN models.}
    \label{fig:bops_and_nabs}
\end{figure}
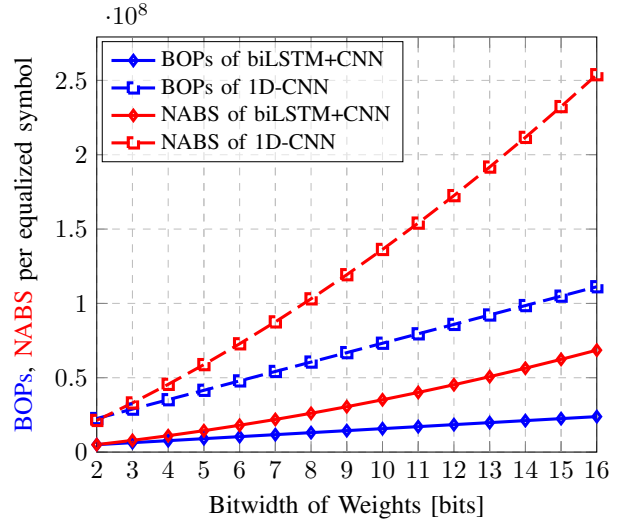

Table.~\ref{tab:KD_models} shows the summary of the performance versus complexity of these two NN architectures when the precision is defaulted from Tensorflow. The bitwidth of the input (b$_i$) is 64 bits, and the bitwidth of the weights (b$_w$) and activation function (b$_a$) is 32 bits. It can be seen that the Q-factor of the student model is slightly lower than that of the teacher model when trained with the simulated data mentioned above\footnote{Note that in the experiment, this KD approach did not show the performance degradation in the student model\cite{srivallapanondh2023parallelization}.}. Even though the number of trainable parameters, RM, BOPs, and NABS of the student model are also higher than the teacher model, the inference latency of the student is actually lower. The inference time analysis was carried out by using CPU (Intel Xeon Processor 2.20 GHz) and GPU (Tesla T4) on Google Colab\cite{voon2021performance}\footnote{Note that, in this study, we did not consider the GPU-accelerated library of primitives for deep neural networks cuDNN (NVIDIA CUDA\textsuperscript{\textregistered}}. This highlights the importance of the ``parallelization'' strategy, which helps reduce the complexity of the processing at the hardware synthesis level. This is crucial for real-world implementation. For the training phase in this case, the trainable parameters can indicate the complexity to some extent; however, empirically, the biLSTM+CNN has a longer training time, even though the number of trainable parameters is lower than the 1D-CNN model. This occurs because the recurrent structure of the biLSTM prevents the computation from being fully parallelizable. At each time step of the calculation, the recurrent structure takes into account the output of the previous time step. This sequential nature makes the training and inference longer. 

Fig.~\ref{fig:bops_and_nabs} shows BOPs and NABs as a function of the bitwidth of the weights (b$_w$) of the NN. It can be observed that the NABS grows with a steeper slope than the BOPs when the b$_w$ increases. This fact highlights that towards the implementation of resource-constrained devices or hardware accelerators, both metrics should be considered carefully, because if only BOPs is assessed at higher precision of the weights, while BOPs fits the requirement, NABS which escalates faster might exceed the requirement of the implementation.

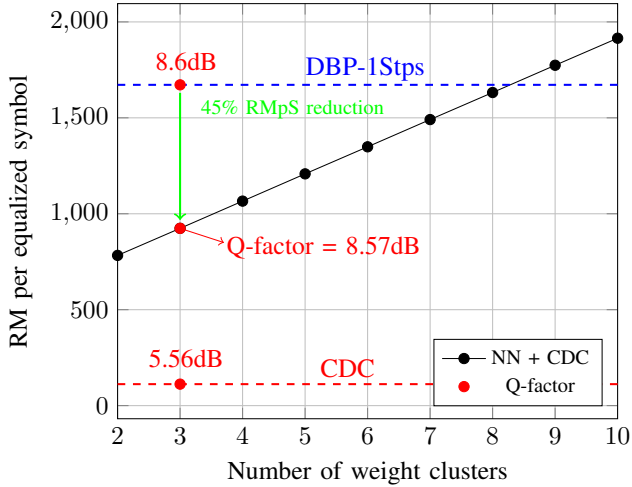
\begin{figure}[ht!]
    \centering
\begin{tikzpicture}[scale=0.97]
    \begin{axis}[
        xlabel={Number of weight clusters},
        ylabel={RM per equalized symbol},
        legend style={font=\footnotesize},
        legend pos=south east,
        grid=both,
        xtick={2,3,4,5,6,7,8,9,10},
        xmin=2,
        xmax=10
    ]
    \addplot[mark=*] coordinates {
    (2, 783)
    (3, 924)
    (4, 1066)
    (5, 1208)
    (6, 1349)
    (7, 1491)
    (8, 1632)
    (9, 1774)
    (10, 1915)
    };
    \addlegendentry{NN + CDC}
    
    \addplot[mark=*, color=red, only marks] coordinates {
    (3, 112)
    (3, 924)
    (3, 1672)
    };
    \addlegendentry{Q-factor}
    
    \addplot[mark=none,color=red,dashed,thick] coordinates {
        (2,112)
        (10,112)
    };

    \addplot[mark=none,color=blue,dashed,thick] coordinates {
        (2,1672)
        (10,1672)
    };

    

    \draw[->, red] (axis cs:3, 924) -- (axis cs:3.7, 850);
    \node[align=center, anchor=south, red] at (axis cs:5.3, 720) {Q-factor = 8.57dB};

    \node[align=center, anchor=south, red] at (axis cs:3.1, 150) {5.56dB};
    \node[align=center, anchor=south, red] at (axis cs:3.1, 1700) {8.6dB};
    \draw[->, green,thick] (axis cs:3, 1630) -- (axis cs:3, 967);
    \node[align=center, anchor=south,  green, font=\footnotesize] at (axis cs:4.8 , 1450) {45\% RMpS reduction};
    \end{axis}
    \node[text width=4.5cm, blue] at (4.9,4.8) 
    {DBP-1Stps};

    \node[text width=4.5cm, red] at (5.1,0.7) 
    {CDC};
\end{tikzpicture}

    \caption{Comparison of complexity (RMpS) between  bLSTM+CNN and traditional channel equalizers (DBP 1STpS and CDC) as a function of the number of clusters utilized in weight clustered compression. The Q-factor in dB is emphasized for the scenario involving 3 clusters. }
    \label{fig:finalcompelxity}
\end{figure}

Finally, it is imperative to address whether the current set of complexity reduction techniques suffices to render NN-based equalizers an appealing choice for industrial implementation. To explore this, we conducted a comparative analysis of the RMpS between traditional DBP 1STpS and CDC implemented in the frequency domain. The results are depicted in Figure~\ref{fig:finalcompelxity}.

To maintain complexity constraints, we employed the same teacher NN architectural with an adaptation involving 50 hidden units in the LSTM layer. This adjustment yielded a Q-factor performance of 8.57dB, comparable to DBP 1STpS (8.6 dB), and 3 dB higher than CDC performance (5.56dB) for the same experiment previously discussed.

In terms of RMpS, while achieving similar performance levels, the complexity of the NN-equalizer + CDC was reduced by 45\% compared to DBP 1STpS when considering 3 weight clusters. However, upon comparing the complexity of the NN equalizer with CDC, it becomes evident that further strides are necessary to achieve lower complexity levels, particularly in terms of multiplication operations. Notably, while CDC typically requires fewer than 100 multiplications per recovered symbol, the NN, even after compression, necessitated between 500-1000 multipliers in its lower complexity configurations. In contrast, the DBP typically demands more than 1500 multipliers per recovered symbol. This underscores the ongoing challenge to optimize the NN-equalizer for reduced complexity, especially in the realm of multiplicative operations, and further investigations are still needed.

\section{Conclusion} \label{sec:open_challenges}

In this study, we explore various methods for creating neural network equalizers with lower complexity in the training, inference, and hardware synthesis stages. To gauge the effectiveness of these techniques, we introduce primary metrics for evaluating NN complexity during both training and inference. During training, factors such as the number of trainable parameters, total training time, the product of epochs and batches, data parallelism, and the number of operational ranges should be considered. Moving to the inference phase,  which is the heart of the real-time operation, we proposed a detailed breakdown of computational complexity into three hardware-agnostic measures: number of real multiplications (RM), number of bit operations (BOP), and number of additions and bit shifts (NABS). This granular approach provides a clear picture of complexity as we transition from software to hardware levels. Notably, the fourth metric, the number of logic gates, is hardware-dependent and requires specific setup information for calculation. The introduction of these detailed metrics allows us to establish a consistent baseline for complexity calculation, catering to varied purposes. We investigate the computation of RM, BOP, and NABS across various NN layers, including dense layers, 1D-convolutional layers, vanilla RNNs, LSTMs, GRUs, and ESN architectures in a general form. 

Our evaluation of the RM metric shows how complexity evolves with changes in different hyperparameters for each NN layer. Notably, LSTM exhibits the highest complexity among recurrent layers due to its architecture featuring different gates, with complexity escalating significantly with increased hidden units. Vanilla RNN emerges as the least complex recurrent architecture. For all recurrent networks, the most impactful hyperparameter on complexity is the number of hidden units, followed by the number of time steps, while the size of the input vector has the least influence. The dense layer stands out as the most economical in complexity due to its simple matrix multiplication.

We also emphasize the importance of bitwidth (precision) in defining BOP complexity as we move closer to the hardware level. A two-fold reduction in bitwidth, particularly in weights, drastically reduces BOP by around 40\% across all types of NNs. This underscores the recommendation to prioritize low-precision bits in NN weights for a more significant reduction in complexity.

Additionally, we introduce the NABS metric to highlight the effects of different quantization techniques (uniform, PoT, and APoT). Unlike other studies claiming drastic complexity reduction with PoT, our work using NABS shows that the true complexity reduction with PoT is only around one order of magnitude compared to uniform quantization for all NN layers. We argue that NABS is a more accurate metric for identifying true complexity levels compared to previously used metrics like RM or BOP.

Finally, a practical study investigates the impact of complexity reduction methods on equalization performance. Our findings demonstrate that implementing these strategies results in NN models with reduced complexity while maintaining a Q-factor level similar to the original counterparts, showcasing the feasibility of complexity reduction for practical applications.

\Urlmuskip=0mu plus 1mu\relax
\bibliographystyle{IEEEtran}
\bibliography{references}

\end{document}